\documentclass[letterpaper,english,reprint, aps, showpacs]{revtex4-1}
\usepackage[T1]{fontenc}
\usepackage[latin9]{inputenc}
\usepackage[active]{srcltx}
\usepackage{xcolor}
\usepackage{array}
\usepackage{float}
\usepackage{textcomp}
\usepackage{multirow}
\usepackage{amsmath}
\usepackage{amssymb}
\usepackage{cancel}
\usepackage{graphicx}
\PassOptionsToPackage{normalem}{ulem}
\usepackage{ulem}

\makeatletter

\providecommand{\tabularnewline}{\\}
\providecolor{lyxadded}{rgb}{0,0,1}
\providecolor{lyxdeleted}{rgb}{1,0,0}

\DeclareRobustCommand{\lyxsout}[1]{\ifx\\#1\else\sout{#1}\fi}

\makeatother

\usepackage{babel}
\begin{document}
\title{Single Production of Vectorlike Y Quarks at the HL-LHC}
\author{V. Cetinkaya}
\address{Department of Physics, Kutahya Dumlupinar University, 43100 Kutahya,
Turkey}
\author{A. Ozansoy}
\address{Department of Physics, Ankara University, 06100 Ankara, Turkey}
\author{V. Ari}
\address{Department of Physics, Ankara University, 06100 Ankara, Turkey}
\author{O. M. Ozsimsek}
\address{Department of Physics Engineering, Hacettepe University, 06800 Ankara,
Turkey}
\author{O. Cakir}
\affiliation{Department of Physics, Ankara University; 06100 Ankara, Turkey}
\begin{abstract}
We study single production of exotic vectorlike $Y$ quark with electric
charge $|Q_{Y}|=4/3$ and its subsequent decay at the High Luminosity
LHC (HL-LHC). Most of the vector like quark (VLQ) decays have the
electroweak $W$ bosons in the intermediate state. Besides their direct
productions singly or pairs, the $W$-bosons are involved in decay
chains as a result of the decay of a top quark which contributes to
the background. This is particularly the case since vectorlike $Y$
quark, which is estimated to be produced with a high cross-section,
can only decay via a $W$ boson and a down type quark ($d,s,b$).
We calculate the cross sections of signal (for different couplings
and mass values) and relevant Standard Model (SM) backgrounds. After
a fast simulation of signal and background events, estimations of
the sensitivity to the parameters (mass range 1000-2500 GeV for coupling
value $\kappa_{Y}=0.5$, and mass range 500-2000 GeV for coupling
values $\kappa_{Y}=0.3$ and $\kappa_{Y}=0.15$) have been presented
at the HL-LHC with center of mass energy $\sqrt{s}=14$ TeV and integrated
luminosity projections of 300 fb$^{-1}$, 1000 fb$^{-1}$ and 3000
fb$^{-1}$.
\end{abstract}
\maketitle

\section{Introduction}

The results from the experiments at the Large Hadron Collider (LHC)
have confirmed the validity of the Standard Model (SM) of particle
physics up to a high energy scale and intensity. The upgrade of the
LHC to the high luminosity phase (HL-LHC) \citep{key-1} at center-of-mass
energy of 14 TeV and integrated luminosity of 3000 fb$^{-1}$ will
extend the sensitivity and perspectives, with the upgraded detectors
and large data, to possible opportunities beyond the SM. This upgrade
will be crucial for precision measurements in the Higgs sector and
for increasing the new physics discovery potential at the energy and
luminosity frontier. The large data will be collected in two steps.
During the first upgrade phase (Run-3), experiments are expected to
collect an integrated luminosity of 300 fb$^{-1}$, whereas in the
next phase (Run-4), a total amount of data corresponding to 3000 fb$^{-1}$
is foreseen.

Predictions for the existence of new fermionic resonances referred
to as vectorlike quarks, which are also common in some beyond the
Standard Model (BSM) scenarios, have been expressed recently. Vectorlike
quarks \citep{key-2,key-3} are defined as colour-triplet under $SU(3)$
and spin-1/2 fermions whose left-handed and right-handed chiral components
have the same transformation properties under the $SU(2)\times U(1)$
gauge group.

The ATLAS \citep{key-4} and CMS \citep{key-5} Collaborations have
published searches for single production of vectorlike $T/Y$ quarks
in decay channel $T/Y\to Wb$ and set $95\%$ confidence level (C.L.)
lower limits on $T$/$Y$ quark masses. The upper limits on the couplings
are $|\sin \theta_{L}|=0.18$ for a singlet $T$ quark, and $|\sin \theta_{R}|=0.17$
for a $(B,Y)$ doublet model, and $|\sin \theta_{L}|=0.16$ for a $(T,B,Y)$
triplet model for a $Y$ quark mass of $800$ GeV. Within the $(B,Y)$
doublet model, the limits on the mixing parameter is comparable with
the exclusion limits from electroweak precision observables in the
mass range {[}$900-1250${]} GeV \citep{key-4}. Upper limits are
placed on the production cross section of heavy exotic quarks by the
CMS experiment \citep{key-5}, for $Y$ quarks with coupling of $0.5$
and $B(Y\to bW)=100\%,$ the observed (expected) lower mass limits
are given as $1400$ ($1000$) GeV.

Potential of the HL-LHC in searching
for the bounds on mass and coupling would be higher. 
In the analysis, we investigate a resonance particle as vectorlike
$Y$ quark in the invariant mass distributions. In order to measure its charge,
leptonic decay of $W$ boson and the charge of $b$-tagged jet can be used.
The single production of vector like quarks is model dependent, 
the framework of the model suggests vectorlike $Y(-4/3)$ quark to 
$Wb$ channel with a branching ratio of $100\%$, however 
vectorlike $T(2/3)$ quark decay into the same channel 
but with different branching (for example $50\%$ for singlet configuration).
This will effect the cross section times branching ratio and one expect stronger bounds 
for the $Y$. Usually experimental searches for vectorlike quarks 
adopt a phenomenological approach, assuming that only one 
new VLQ state is present beyond the SM.
The analysis has been designed based on a simplified 
scenario for modelling the VLQ dynamics, assuming that only one 
new VLQ is present beyond the SM for parametrising 
its single production. Single production of vectorlike quarks with
large width effects have already been studied in Refs. \citep{key-6,key-7}.

In this study, we consider an effective model framework for the single
production of vectorlike $Y$ quark in the second section. In the
third section, we mention about the decay width of vectorlike $Y$
quark.
Production cross sections for signal process as well as corresponding
SM backgrounds are given in the fourth section. Modeling of the signal
and background events are performed in the fifth section. After detector
simulation, event selection and analysis results have been presented
in the sixth section. Finally, statistical significance of the signal
have been given depending on the parameter space (mass and coupling)
of the model framework, and we draw a conclusion on the search potential
for vectorlike $Y$ quark at the HL-LHC.

\section{Model Framework}

Depending on the model framework, vectorlike quarks are classified
as $SU(2)$ singlets, doublets or triplets of flavours $B,T,X$ or
$Y$, in which the first two have the same charge as the standard
model (SM) bottom and top quarks while the vectorlike quark $X$ and
$Y$ have exotic electric charge $5e/3$ and $-4e/3$, respectively.
In this framework, vectorlike quark $Y$ can exists as ($B,Y$) doublet
or ($T,B,Y$) triplet. As in the description detailed in Ref. \citep{key-8},
an exotic vectorlike $Y$ quark can decay into a $W$ boson and a
down sector ($d,s,b$) quark. Within the framework, vectorlike quarks
are expected to couple preferentially to third-generation quarks and
can have flavour-changing neutral-current decays in addition to the
charged-current decays characteristic of chiral quarks \citep{key-8}.
We use an effective Lagrangian framework for the interactions of vectorlike
$Y$ quark with the SM quarks through the $W$ boson exchange including
free parameters ($\kappa_{qL}^{Y},\kappa_{qR}^{Y}$):

\begin{align}
L_{Y} & =i\bar{Y}\!\!\cancel{D}Y\text{\textminus}m_{Y}\bar{Y}Y\nonumber \\
 & +\frac{g}{\sqrt{2}}[\bar{Y}\!\!\cancel{W}(\kappa_{dL}^{Y}P_{L}+\kappa_{dR}^{Y}P_{R})d+\text{h.c.}\nonumber \\
 & +\frac{g}{\sqrt{2}}[\bar{Y}\!\!\cancel{W}(\kappa_{sL}^{Y}P_{L}+\kappa_{sR}^{Y}P_{R})s+\text{h.c.}\nonumber \\
 & +\frac{g}{\sqrt{2}}[\bar{Y}\!\!\cancel{W}(\kappa_{bL}^{Y}P_{L}+\kappa_{bR}^{Y}P_{R})b+\text{h.c.}\label{eq:1}
\end{align}
where $\cancel{D}$ and $\cancel{W}$ are the definitions $\gamma^{\mu}D_{\mu}$
and $\gamma^{\mu}W_{\mu}$ for the covariant derivative and $W$ boson
field, respectively \citep{key-9}. The terms in the first line denote
gauge invariant kinetic and mass term for the vectorlike $Y$ quark
fields, both strong and electroweak pieces of the covariant derivative
can be included in this part. However, the electroweak pieces are
expected to yield a small effect with respect to their strong interaction
part \citep{key-8}. The next three lines define the interactions
between vectorlike $Y$ quark and SM down type quarks ($d,s,b$) through
a $W$ boson exchange. Vectorlike quarks can also induce a mixing
between the SM and new physics (NP) sectors. The corresponding elements
of the mixing matrices can be inserted in the interactions strength.
The $\kappa_{qL}^{Y}$ and $\kappa_{qR}^{Y}$ parameters include the
relevant elements of the quark mixing matrices with the corresponding
projection operators $P_{L}$ and $P_{R}$. The mixing of VLQs simultaneously
with more than one SM family is strongly constrained by various flavor
changing processes and we focus on a mixing with the third family
in the work . Following this framework interaction terms in the Lagrangian
(Eq. \ref{eq:1}) have been supplemented into the SM Lagrangian.

\section{Decay of VectorLike Y Quark}

The decay width of vectorlike $Y$ quark has been expressed as

\begin{equation}
\Gamma(Y\to Wq)=\frac{\alpha_{e}(\kappa_{qL}^{2}+\kappa_{qR}^{2})}{16s_{W}^{2}}\frac{(m_{W}^{2}-m_{Y}^{2})^{2}(2m_{W}^{2}+m_{Y}^{2})}{m_{W}^{2}m_{Y}^{3}}\label{eq:2}
\end{equation}
where down-type SM quarks ($q=d,s,b$) are taken massless. The interaction
strength can be parametrized in terms of electromagnetic coupling
constant $g_{e}=\sqrt{4\pi\alpha_{e}}$ and $s_{W}$ (which stands
for the sine of the electroweak mixing angle). The relative importance
of a decay channel, for instance $\Gamma(Y\to Wq)$, can be expressed
by the ratio $\Gamma(Y\to Wq)/\sum_{q}\Gamma(Y\to Wq)=(\kappa_{qL}^{2}+\kappa_{qR}^{2})/\sum_{q}(\kappa_{qL}^{2}+\kappa_{qR}^{2})$.
Taking into account the current limits on the mass and couplings of
vectorlike quarks from high energy experiments, and sensitivity to
the new physics parameters, we focus on that vectorlike $Y$ quark
couples only to the bottom quark through the charged current interacton
($\kappa_{Y}=\kappa_{bL}$ or $\kappa_{bR}$). For large mass values
of vectorlike $Y$ quark, i.e. ratio $m_{W}/m_{Y}\ll1$, the decay
width is approximated as $\Gamma(Y\to Wb)\sim A\kappa_{Y}^{2}m_{Y}^{3}$,
where $A$ is a constant with a value of $3.28\times10^{-7}$ GeV$^{-2}$.

\begin{figure}
\includegraphics[scale=0.45]{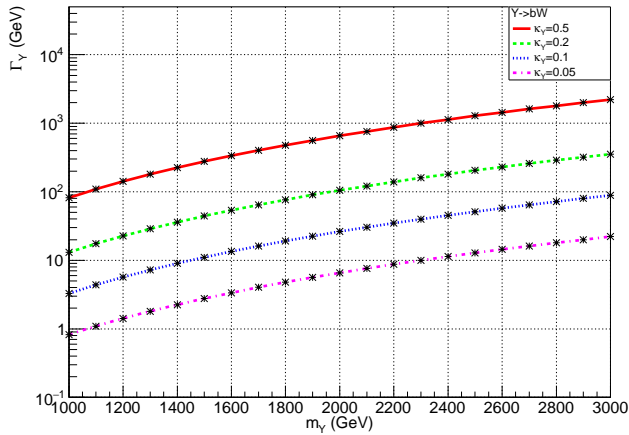}

\caption{Decay width for vectorlike $Y$ quark depending on the mass $m_{Y}$
for different value of the coupling parameter $\kappa_{Y}$.
\label{fig:fig1}}
\end{figure}

The decay width for vectorlike $Y$ quark depending on the mass $m_{Y}$
for different value of the coupling parameter $\kappa_{Y}=0.5,0.2,0.1,0.05$
has been presented in
Fig. \ref{fig:fig1}. As it can be seen from this figure, decay width
changes quadratically with the values of coupling parameters and changes
as a function of mass approximately ($\approx m_{Y}^{3}$) as given
in Eq. \ref{eq:2}. Given the specified width, the couplings are implicit
in the model for fixed mass value.

\section{Production Cross Section}

\begin{figure}
\includegraphics[scale=0.4]{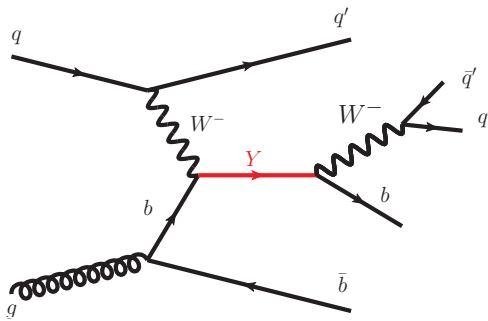}

\caption{Representative diagram for single production of vectorlike $Y$ quark
with subsequent hadronic decay channel. \label{fig:fig2}}
\end{figure}

In order to make a prediction for the signal, we calculate cross section
for on-shell vectorlike $Y$ quark production. Here, we consider $pp\to Ybj+X$
signal process for investigating interactions between vectorlike $Y$
quark and SM bottom quark via $W$ boson specified in Eq. \ref{eq:1}.
The representative diagram for the subprocess $qg\to Y\bar{b}q'$
(a similar diagram for $\bar{Y}$ single production) with subsequent
decays $Y\to W^{-}b\to q\bar{q}'b$ is presented in Fig. \ref{fig:fig2}.
It is representative for the brevity, because there are a lot of signal
diagrams for both $Y\bar{b}j$ and $\bar{Y}bj$ productions, corresponding
to the $W^{\pm}$ exchange, light quark flavors $q$ and $q'$, \textbf{$b$}-quark
and anti $b$-quark, which we take them into account in the event
generation.

\begin{figure}
\includegraphics[scale=0.45]{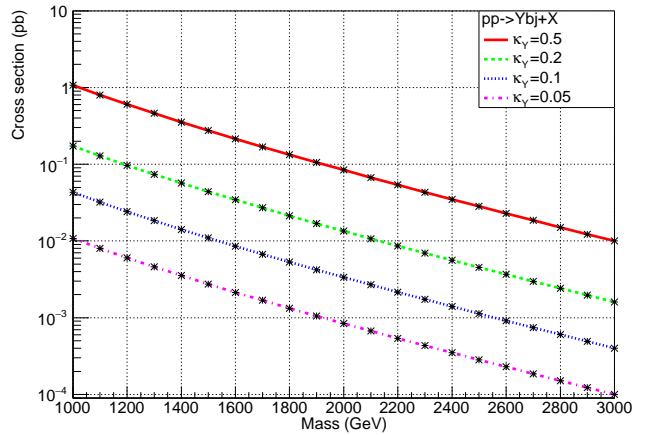}

\caption{Cross section for the process $pp\to Ybj+X$ depending on the mass
of vectorlike $Y$ quark for different coupling values. \label{fig:fig3}}
\end{figure}

The cross section for single production of vectorlike $Y$ quark through
the process $pp\to Ybj+X$ depending on different mass $m_{Y}$ and
coupling values $\kappa_{Y}$ has been presented in Fig. \ref{fig:fig3}.
As it can be seen from this figure, the cross section has large values
especially in the low mass region and it increases depending on the
increase in coupling parameters. In more detail, we present the signal
cross section for different mass $m_{Y}=1000,1500,2000,2500$ GeV
and coupling parameter values $\kappa_{Y}=0.5,0.2,0.1,0.05$
in Table \ref{tab:tableA1}. We present cross section numerical values
in the Appendix A to avoid similar information as in Fig. \ref{fig:fig3}. The cross section decreases according to the decreasing
values of coupling parameter $\kappa_{Y}$ and to the increasing values
of the mass $m_{Y}$. While making these cross section calculations,
we take into account automatic calculation (auto) mode of decay width
$\Gamma_{Y}$. In the single production of vectorlike $Y$ quark,
which is model dependent, the cross section can be related to a function 
of coupling $\kappa_{Y}$ and mass $m_{Y}$ within the model framework.

\section{Modeling of Signal and Background}

The signal samples are generated with the MadGraph5\_aMC@NLO \citep{key-10}
using VLQ\_V4\_4FNS\_UFO model implemented in FeynRules \citep{key-11,key-12},
through single production $pp\to Ybj$ as a high mass on-shell particle
$Y$ with left-handed or right-handed couplings and subsequent decays
$Y\to Wb$ followed by $W$ boson decaying hadronically. However,
the kinematics of the final-state jets are similar for left-handed
($\kappa_{bL}^{Y}$) and right-handed ($\kappa_{bR}^{Y}$) couplings,
then the acceptances for two chiralities are found to be the same.
All hadronic channel is optimised to search for massive vectorlike
$Y$ quark which decays to a $W$ boson and a high-momentum $b$-jet
in the final state. The $W$-boson and the $b$-jet originating from
the vectorlike $Y$ quark decay are expected to be back-to-back in
the transverse plane. The signal topology includes an outgoing light
quark in the process which often produces a forward jet in the detector.
The second $b$-jet in the signal events comes from the gluon to a
pair of $b$-quark ($b\bar{b}$) splitting which may be observed in
either the forward or central region. Having typically low momentum,
it is often assumed to be outside of the detector acceptance.

Signal and background cross sections are given in Table \ref{tab:table1}.
The SM background simulation samples include top quark pair $t\bar{t}$,
$W+jets$, $Z+jets$, single top quark ($tj,$ $tb$, $tW$), associate
top and $Z$ production ($tZj$), $WW$, $ZZ$ and $WZ$ dibosons.
The events for these processes are generated with MadGraph5\_aMC@NLO
\citep{key-10}. Monte Carlo event simulations for SM background and
signal are interfaced with Pythia8 \citep{key-13} for fragmentation
and showering. Signal and background samples use the NNPDF2.3 PDF
set \citep{key-14}. Both the signal and background events are produced
with the generator level cuts: such as minimum transverse momentum
of jets $p_{T}(j)=20$ GeV, maximum pseudo-rapidity for jets $|\eta(j)|=5$,
and minimum distance between jets $\Delta R(j,j)=0.4$.

\begin{table}
\tiny\caption{Process, mode and generated jets in modeling the signals and different
backgrounds, and calculated cross sections. While calculating these
cross sections, the $W/Z$ boson and top quark are allowed to decay
hadronically. \label{tab:table1}}

\begin{tabular}{|c|c|c|c|}
\hline 
Process & Mode & Gen jets & Cross Section (pb)\tabularnewline
\hline 
\hline 
$\kappa_{Y}=0.5$, $m_{Y}=1000\text{ GeV}$ & $Ybj$ & $2b+3j$ & $6.499\times10^{-1}$\tabularnewline
\hline 
$\kappa_{Y}=0.5$, $m_{Y}=1500\text{ GeV}$ & $Ybj$ & $2b+3j$ & $1.520\times10^{-1}$\tabularnewline
\hline 
$\kappa_{Y}=0.5$, $m_{Y}=2000\text{ GeV}$ & $Ybj$ & $2b+3j$ & $4.160\times10^{-2}$\tabularnewline
\hline 
$\kappa_{Y}=0.5$, $m_{Y}=2500\text{ GeV}$ & $Ybj$ & $2b+3j$ & $1.236\times10^{-2}$\tabularnewline
\hline 
$\kappa_{Y}=0.3$, $m_{Y}=500\text{ GeV}$ & $Ybj$ & $2b+3j$ & $1.422\times10^{0}$\tabularnewline
\hline 
$\kappa_{Y}=0.3$, $m_{Y}=1000\text{ GeV}$ & $Ybj$ & $2b+3j$ & $2.367\times10^{-1}$\tabularnewline
\hline 
$\kappa_{Y}=0.3$, $m_{Y}=1500\text{ GeV}$ & $Ybj$ & $2b+3j$ & $5.909\times10^{-2}$\tabularnewline
\hline 
$\kappa_{Y}=0.3$, $m_{Y}=2000\text{ GeV}$ & $Ybj$ & $2b+3j$ & $1.734\times10^{-2}$\tabularnewline
\hline 
$\kappa_{Y}=0.15$, $m_{Y}=500\text{ GeV}$ & $Ybj$ & $2b+3j$ & $3.531\times10^{-1}$\tabularnewline
\hline 
$\kappa_{Y}=0.15$, $m_{Y}=1000\text{ GeV}$ & $Ybj$ & $2b+3j$ & $5.931\times10^{-2}$\tabularnewline
\hline 
$\kappa_{Y}=0.15$, $m_{Y}=1500\text{ GeV}$ & $Ybj$ & $2b+3j$ & $1.513\times10^{-2}$\tabularnewline
\hline 
$\kappa_{Y}=0.15$, $m_{Y}=2000\text{ GeV}$ & $Ybj$ & $2b+3j$ & $4.583\times10^{-3}$\tabularnewline
\hline 
$t\bar{t}$ & $t\bar{t}$ & $2b+4j$ & $2.174\times10^{2}$\tabularnewline
\hline 
\multirow{3}{*}{$SingleTop$} & $tj$ & $b+3j$ & $1.346\times10^{2}$\tabularnewline
\cline{2-4} \cline{3-4} \cline{4-4} 
 & $tb$ & $2b+2j$ & $4.436\times10^{0}$\tabularnewline
\cline{2-4} \cline{3-4} \cline{4-4} 
 & $tW$ & $b+4j$ & $9.267\times10^{-2}$\tabularnewline
\hline 
$tZj$ & $tZj$ & $b+5j$ & $1.040\times10^{-2}$\tabularnewline
\hline 
\multirow{3}{*}{$Dibosons$} & $WW$ & $4j$ & $2.775\times10^{1}$\tabularnewline
\cline{2-4} \cline{3-4} \cline{4-4} 
 & $WZ$ & $4j$ & $8.341\times10^{0}$\tabularnewline
\cline{2-4} \cline{3-4} \cline{4-4} 
 & $ZZ$ & $4j$ & $2.529\times10^{0}$\tabularnewline
\hline 
\multirow{5}{*}{$W+jets$} & $W+j$ & $3j$ & $2.540\times10^{4}$\tabularnewline
\cline{2-4} \cline{3-4} \cline{4-4} 
 & $W+2j$ & $4j$ & $1.062\times10^{4}$\tabularnewline
\cline{2-4} \cline{3-4} \cline{4-4} 
 & $W+3j$ & $5j$ & $4.329\times10^{3}$\tabularnewline
\cline{2-4} \cline{3-4} \cline{4-4} 
 & $W+bj$ & $b+3j$ & $4.564\times10^{0}$\tabularnewline
\cline{2-4} \cline{3-4} \cline{4-4} 
 & $W+bjj$ & $b+4j$ & $1.881\times10^{0}$\tabularnewline
\hline 
\multirow{5}{*}{$Z+jets$} & $Z+j$ & $3j$ & $6.476\times10^{3}$\tabularnewline
\cline{2-4} \cline{3-4} \cline{4-4} 
 & $Z+2j$ & $4j$ & $2.668\times10^{3}$\tabularnewline
\cline{2-4} \cline{3-4} \cline{4-4} 
 & $Z+3j$ & $5j$ & $1.086\times10^{3}$\tabularnewline
\cline{2-4} \cline{3-4} \cline{4-4} 
 & $Z+bj$ & $b+3j$ & $9.010\times10^{-3}$\tabularnewline
\cline{2-4} \cline{3-4} \cline{4-4} 
 & $Z+bjj$ & $b+4j$ & $9.828\times10^{-3}$\tabularnewline
\hline 
\end{tabular}
\end{table}

The signal process $pp\rightarrow Ybj$ (which includes both $Y$
and $\bar{Y}$ VLQs, as well as $b$ and $\bar{b}$ quarks) process
suffers from various backgrounds and one of them is clearly $t\bar{t}$
production process. When considering production and all hadronic decays,
the $t\bar{t}$ process mimics the most characteristic feature of
the signal process which leads the existence of two b-jets $(t\bar{t}\rightarrow W^{+}bW^{-}\bar{b}\to4j+2b)$.

$t\bar{t}$ pair production is an important background in most BSM
searches. Top quark pairs generally decay into two $W$ bosons and
a pair of $b$ quarks. Then, the final state contains at least six
hadronic jets. Moreover due to the high cross section of the $t\bar{t}$
background, it shows presence at every relevant region and resembles
in many aspects of signal process, hence it must be trimmed by applying
proper event selection and analysis cuts. $t\bar{t}$ background and
signal processes differ in some respects such as the existence of
more energetic jets and higher hadronic transverse energy for the
signal. However, presence of two top quarks with high cross section
is obviously problematic. As a direct consequence of this result,
we expect an affection at top mass reconstruction. Hence our cuts
are decided to reduce its dominance.

Single top production also matters and needed to be handled well using
proper cuts. Since it has large cross section and some modes of that
background are very close to signal process besides due to high mass
of top quark, mass reconstruction for these processes resemble in
some aspects with the signal. Single top quark production $tq$, $tb$
and $tW$ samples, in which top quark decays to a $W$ boson (decaying
hadronically) and a $b$ quark ($t\rightarrow Wb\rightarrow q\bar{q'}b$),
have been produced. The $tZj$ sample has been produced with the top
quark decaying hadronically
($t\rightarrow Wb\rightarrow q\bar{q'}b$) and $Z$ boson decaying
to two quarks ($Z$$\rightarrow q\bar{q}$($q\neq b$)).

We produce $W$+$jets$ and $Z$+$jets$ samples which include only
hadronic $W$ or $Z$ boson decays, and a hadronic transverse energy
cut $H_{T}>400$ GeV (where $H_{T}$ is the scalar sum of transverse
momentum of selected jets) is applied in the analysis to reduce these
background.

The dibosons $WW$, $ZZ$ and $WZ$ samples have been generated with
hadronic decays leading at least four hadronic jets, which can be
suppressed according to the selections of two jets invariant mass
interval.

At largely, the backgrounds mentioned after single top are less problematic.
Although some of them have large cross sections, with the invariant
mass reconstruction, cut down on the signal dominant region. Futhermore,
transverse momentum of the leading jet have significantly different
behaviour for the signal. On the whole, background can be kept under
control especially at high mass region.

The detector response to the signal and background events is simulated
using a detailed description of the HL-LHC detector card implementation
in Delphes v3.4 \citep{key-15}. All events are analyzed by developing
an analysis macro with Root v6 \citep{key-16} . The kinematical distributions
are normalized to the number of events, which is defined to be the
cross-section times luminosity for each process under consideration.

\section{Analysis And Results}

The analysis targets $Wb$ events with $W$ boson decaying hadronically
in order to reconstruct vectorlike $Y$ quarks (full hadronic mode:
$Y\to bW\to bjj$). In this mode event selection requires at least
five jets and at least one of them being $b$-tagged, where hadronically
decaying $W$ boson identified from dijet invariant mass. Our all
hadronic analysis focuses on final states at least five (small radius)
jets, one of them can be identified as forward jet and one of the
$b$-tagged jet can be considered collinear jet, other two jets can
be considered as central jets reconstructing the $W$ boson mass.
These two jets together with one leading ($b$-tagged) jet reconstruct
vectorlike $Y$ quark invariant mass. The final discriminant variables
are chosen as leading jet $p_{T}$, hadronic transverse energy $H_{T}$
(We define $H_{T}$ as the hadronic activity typically associated
with single vectorlike $Y$ quark production, where this variable
is defined as the scalar sum of the transverse momentum of selected
jets in the events.), and the reconstructed mass $m_{Y}$ of the vector
like $Y$quark candidate.

\begin{figure}
\includegraphics[scale=0.45]{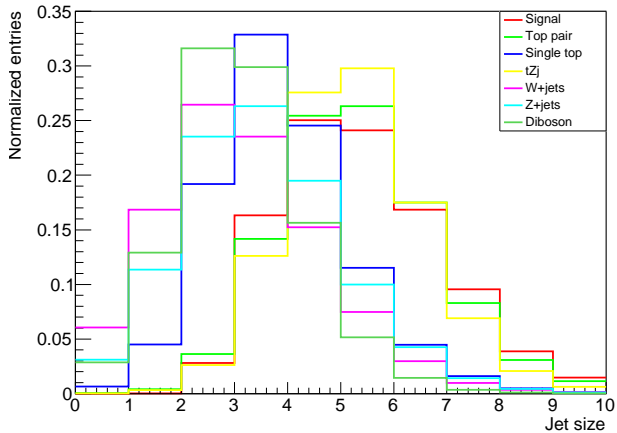}

\caption{The normalized distribution of the number of jets (jet size) for signal
(with $m_{Y}=1500$ GeV and $\kappa_{Y}=0.5$), and also for the relevant
backgrounds in all hadronic channels. \label{fig:fig4}}
\end{figure}

The normalized distributions of the number of jets in signal events
(with mass $m_{Y}=1500$ GeV and coupling $\kappa_{Y}=0.5$), and
also in the relevant backgrounds are given in Fig. \ref{fig:fig4}.
The signal has mostly four and five jets in the final state, while
the background $tZj$ and $t\bar{t}$ dominates in this region, event
selection with the requirement $n_{jet}\geqslant5$ reduce other backgrounds.

One of the important background for our analysis is the top pair production
$t\bar{t}$ due to its high cross section. Here, we give the $p_{T}$
and $\eta$ distributions of the jets ($p_{T}$ ordered) for this
process without any cuts to reflect it's main characteristics. Hence,
we have the chance to exploit its main differences. As we can see
from a quick comparison between the signal distributions, discriminating
differences occur at the distributions of the high $p_{T}$ jets.

The transverse momentum ($p_{T}$) and pseudo-rapidity ($\eta$) distributions
of five jets for signal, for $m_{Y}=1500$ GeV and $m_{Y}=2000$ GeV
($\kappa_{Y}=0.5$ for both) and also for top pair, associated $tZj$
production, single top, $W$+$jets$, $Z$+$jets$ and dibosons background
are presented in Fig. \ref{fig:figA1} - \ref{fig:figA8}, respectively.

The transverse momentum ($p_{T}$) and pseudo-rapidity ($\eta$) distributions
of jets for $W+jets$ (including $W+j,W+2j,W+3j,W+bj,W+bjj$) background
are presented in Fig. \ref{fig:figA6}. The generator level cuts are
applied to these kinematic distributions of the final state jets.
Comparing the signal and the $W+jets$ background distributions, it
can be seen that 1st and 2nd jets have different distributions of
the $p_{T}$ and $\eta$. A similar discussion is also valid for the
$Z+jets$ background when the hadronic mode is considered.

The transverse momentum and pseudo-rapidity distributions of all jets
for whole diboson background (including $WW,WZ,ZZ$) before performing
the cuts are shown in Fig. \ref{fig:figA8}. As it can be seen from
Fig. \ref{fig:figA8}, the pseudo-rapidity distribution of the jets
are symmetrical with respect to the beam axis and the tails of the
distribution for the 5th jet are getting longer in both forward and
backward direction. The transverse momentum distribution falls rapidly
for high $p_{T}$ values.

In addition to the $p_{T}$ and $\eta$ cuts on the selected final
state jets, a transverse hadronic energy cut $H_{T}>400$ GeV is applied
in the analysis, where $H_{T}$ is the scalar sum of the $p_{T}$
of all selected jets. The angular distance (or angular separation)
between two jets defines how much two jets are moving in the same
direction, and it is usually denoted as $\Delta R=\sqrt{\Delta\eta^{2}+\Delta\phi^{2}}$,
where a cut $\Delta R(j,j)>0.4$ is applied to the selected events. The preselection and
cut flow for the analysis are given in Table \ref{tab:table2}.

Cut efficiencies ($\%$) for different signal benchmarks ($\kappa_{Y}=0.5$
and $m_{Y}=1000$, $1500$, $2000,$ $2500$ GeV) and backgrounds
in Table \ref{tab:table3} have been obtained by applying the cuts
explained in Table \ref{tab:table2}, where the leading jet $p_{T}$
variable has been used.

\begin{figure}
\includegraphics[scale=0.4]{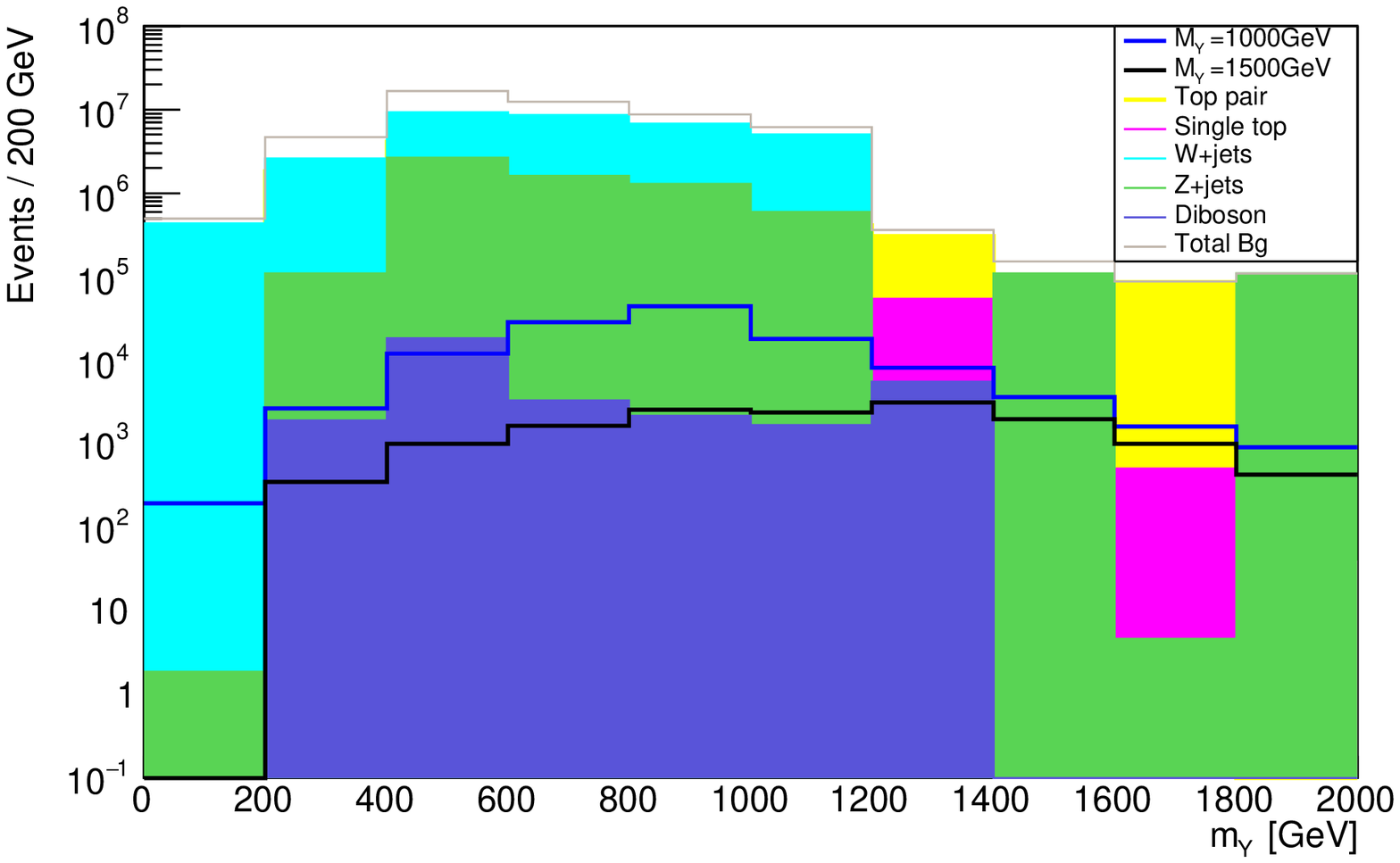}

\includegraphics[scale=0.4]{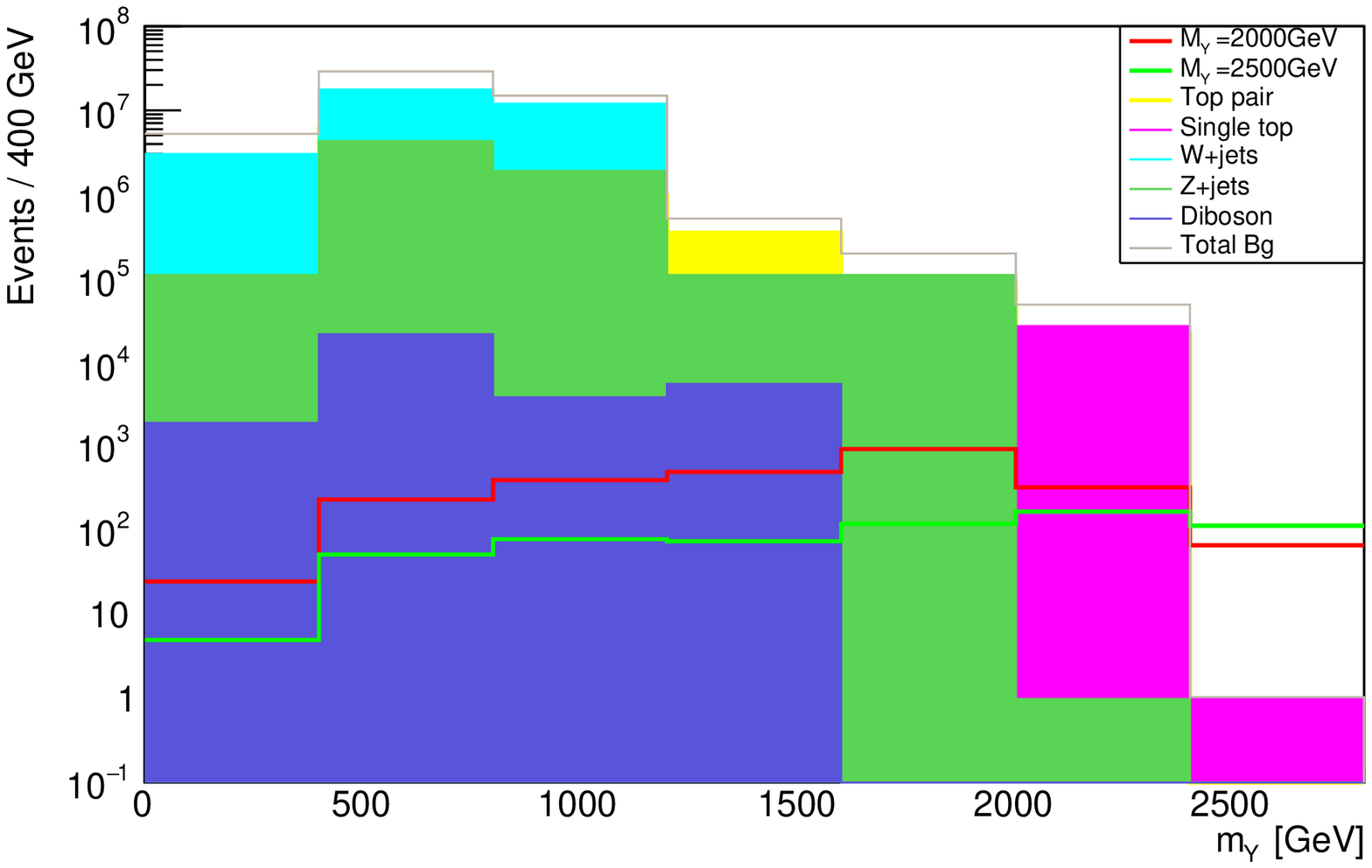}

\caption{The invariant mass distribution $m_{Y}$ for all signals ($\kappa_{Y}=0.5$)
and backgrounds. For $m_{Y}=1000$ and $1500$ GeV (upper), for $m_{Y}=2000$
and $2500$ GeV (lower).\label{fig:fig5}}
\end{figure}

\begin{figure}
\includegraphics[scale=0.45]{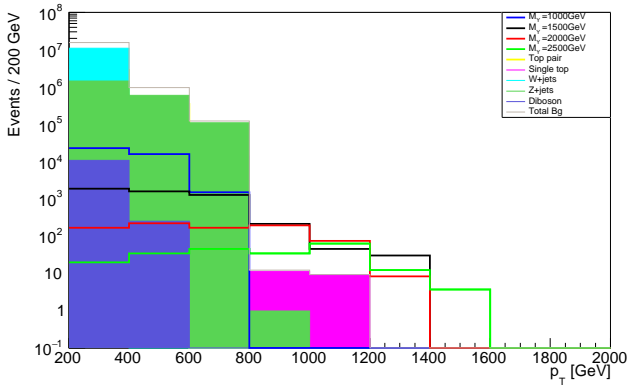}

\caption{The leading jet $p_{T}$ distributions for all signals ($\kappa_{Y}=0.5$)
and backgrounds. \label{fig:fig6}}
\end{figure}

\begin{figure}
\includegraphics[scale=0.45]{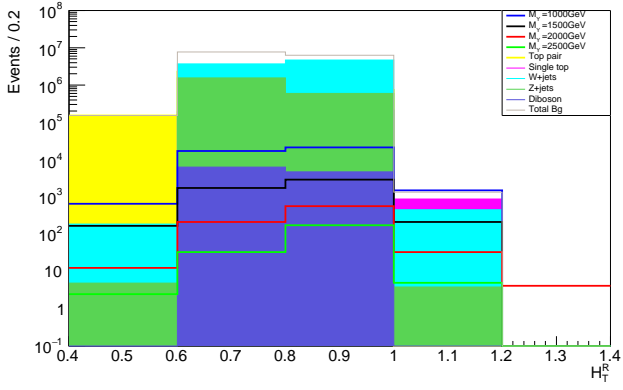}

\caption{The ratio of the scalar sum $p_{T}$ to the total scalar $H_{T}$.
$\kappa_{Y}=0.5$ for all signals.\label{fig:fig7}}
\end{figure}

\begin{figure}
\includegraphics[scale=0.4]{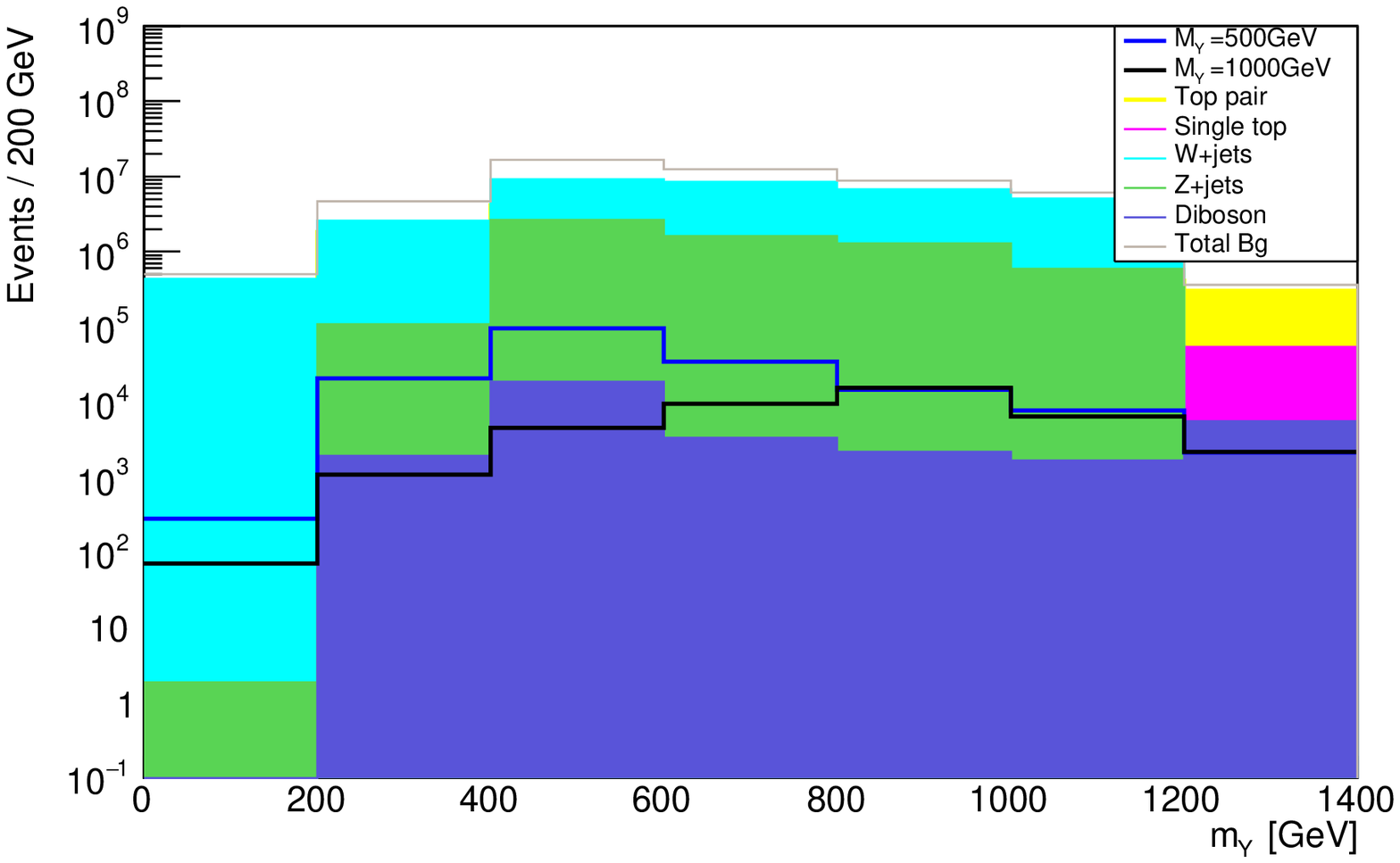}

\includegraphics[scale=0.4]{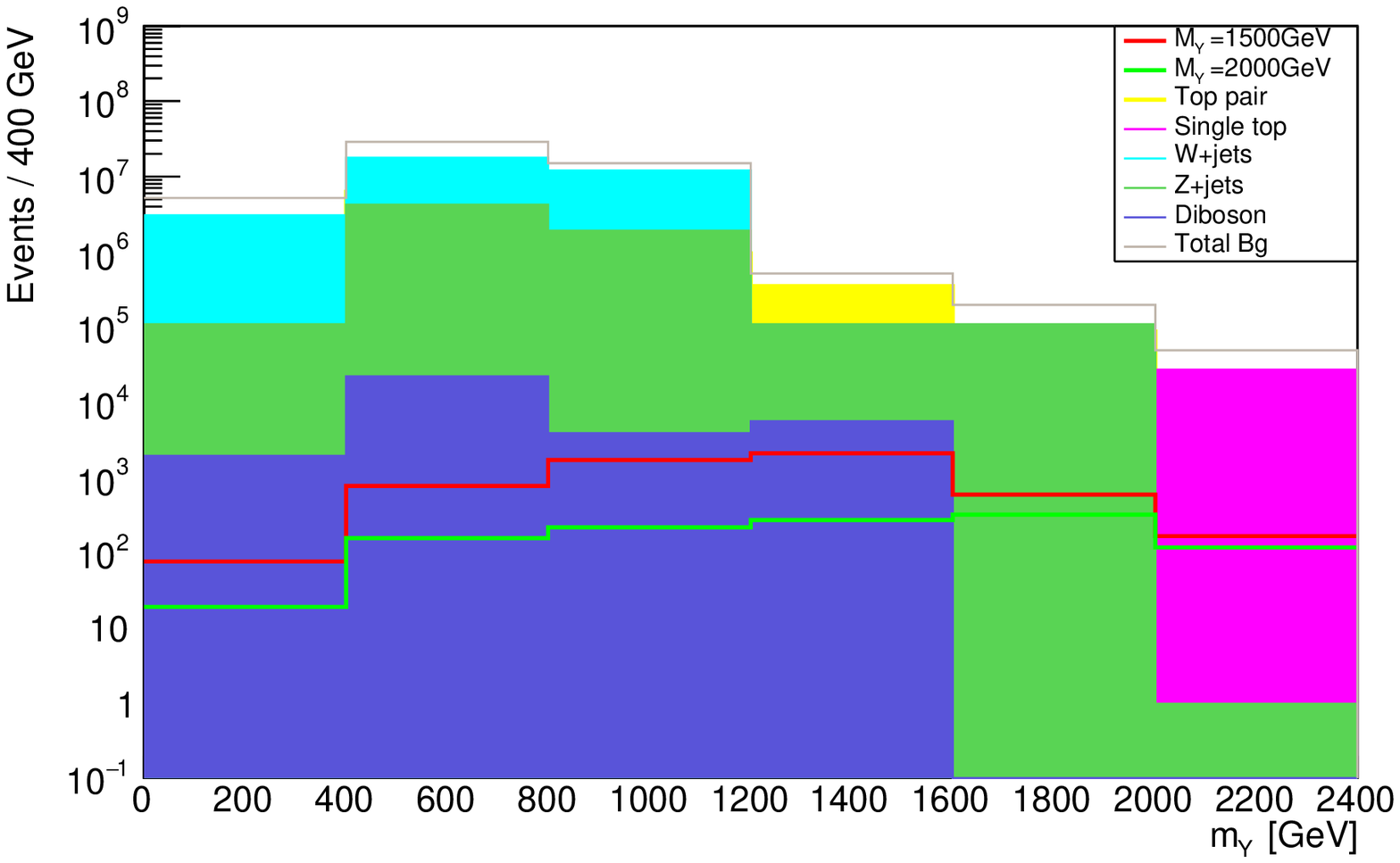}

\caption{The invariant mass distribution $m_{Y}$ for all signals ($\kappa_{Y}=0.3$)
and backgrounds. For $m_{Y}=500$ and $1000$ GeV (upper), for $m_{Y}=1500$
and $2000$ GeV (lower).\label{fig:fig8}}
\end{figure}

\begin{figure}
\includegraphics[scale=0.4]{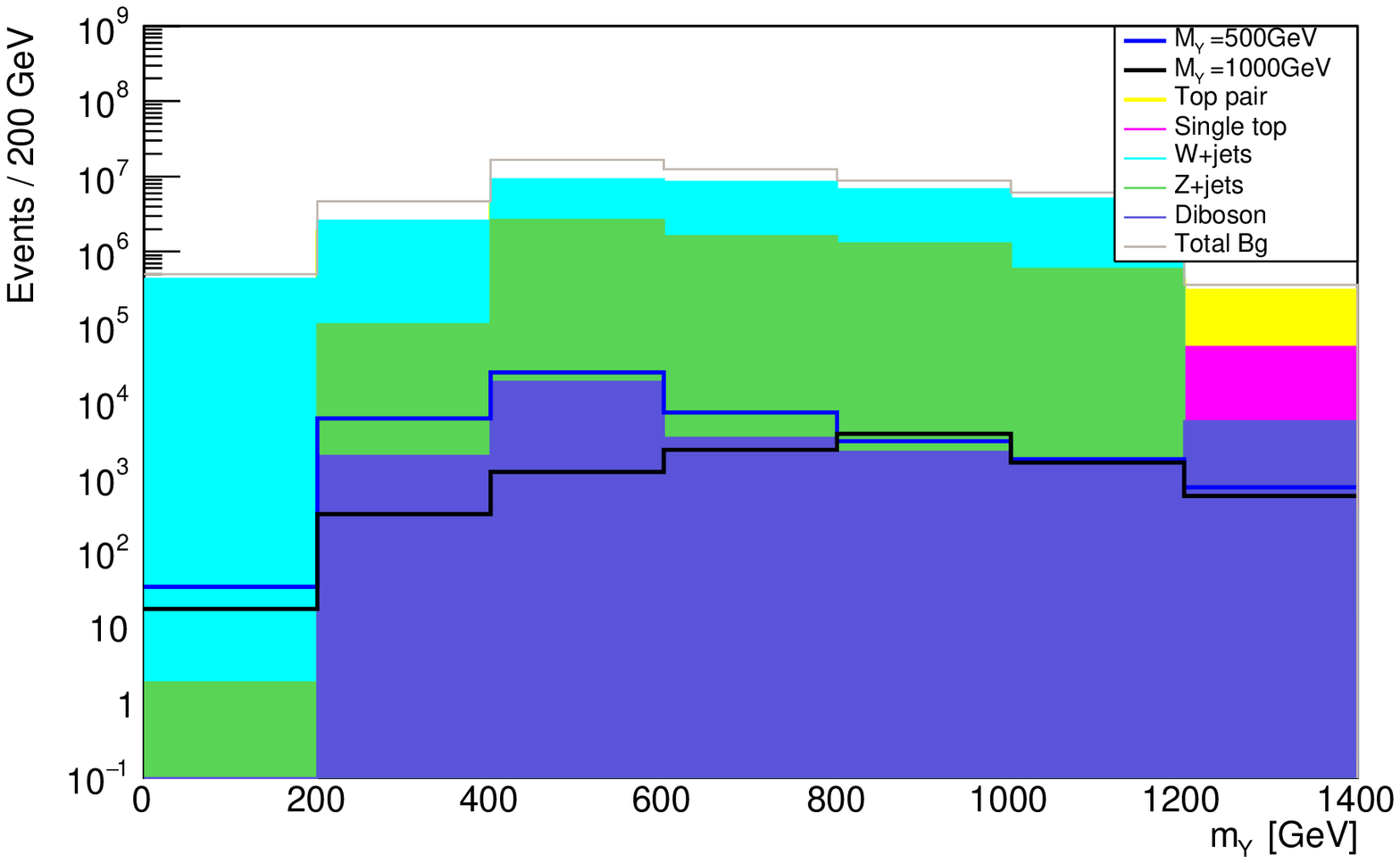}

\includegraphics[scale=0.4]{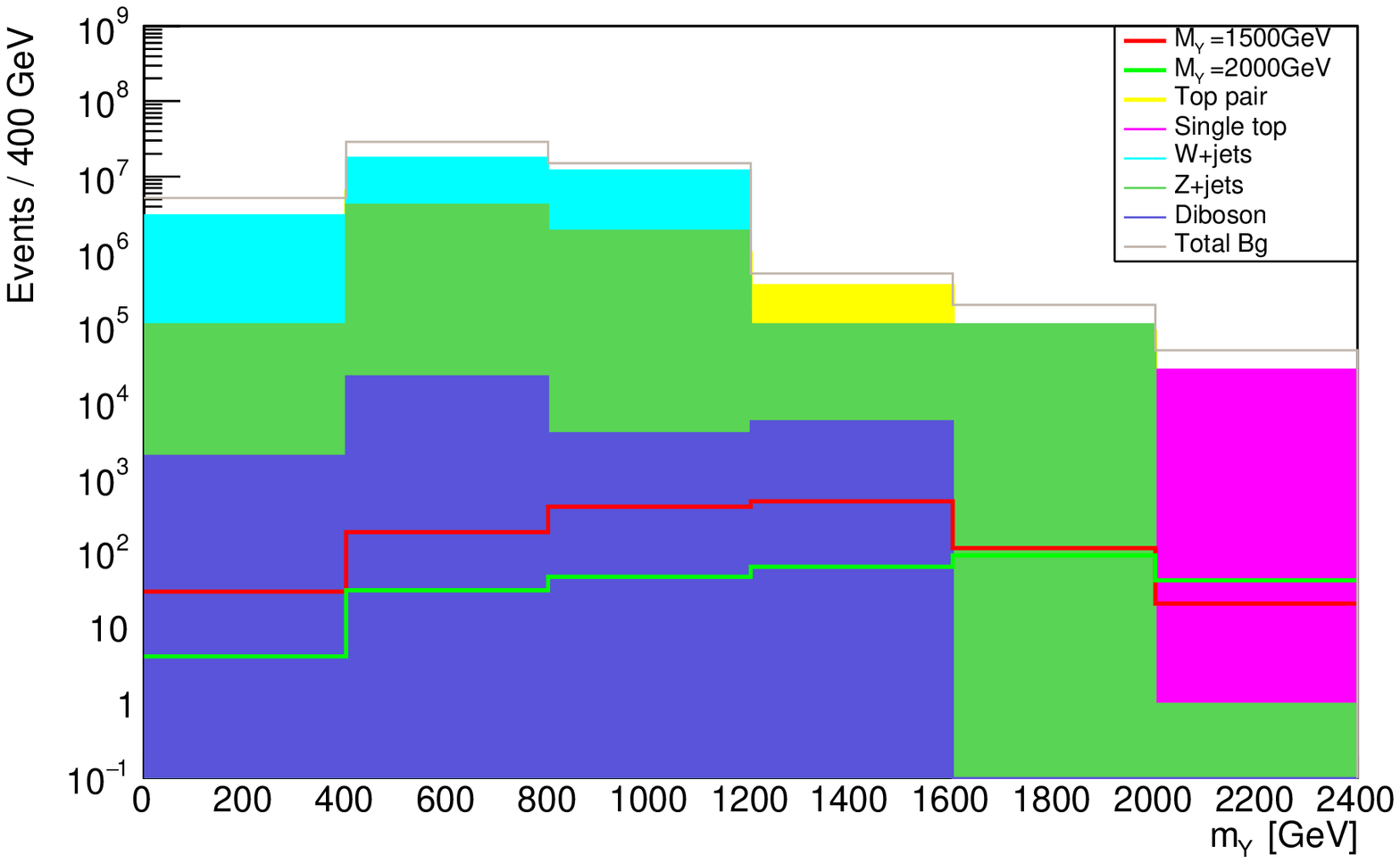}

\caption{The invariant mass distribution $m_{Y}$ for all signals ($\kappa_{Y}=0.15$)
and backgrounds. For $m_{Y}=500$ and $1000$ GeV (upper), for $m_{Y}=1500$
and $2000$ GeV (lower).\label{fig:fig9}}
\end{figure}

The invariant mass distribution $m_{Y}$ for all signals and background
are presented in Fig. \ref{fig:fig5}. The leading jet $p_{T}$ distributions
for all signals and backgrounds are shown in Fig. \ref{fig:fig6}.
The scalar sum $p_{T}$ for all signals and backgrounds are shown
in Fig. \ref{fig:figA9}. The ratio of the scalar sum $p_{T}$ to
the total scalar $H_{T}$, $H_{T}^{R}$, in the event is shown in
Fig. \ref{fig:fig7}. Fig. \ref{fig:fig8} and Fig. \ref{fig:fig9}
show the invariant mass distributions of the vectorlike $Y$quark
for $\kappa_{Y}=0.3$ and $\kappa_{Y}=0.15$, respectively. The distributions
shown in Fig. \ref{fig:fig5}-\ref{fig:fig9} and Fig. \ref{fig:figA9} are normalized to
the expected number of events, defined as the cross-section of related
process multiplied by the integrated luminosity of $L_{int}=1000$
fb$^{-1}$.

\begin{table}
\tiny\caption{Preselection and set of cuts for the analysis of signal and background
events. \label{tab:table2}}

\begin{tabular}{|c|c|}
\hline 
Cuts & Definition\tabularnewline
\hline 
\hline 
Cut-0 & Preselection: number of jets $N_{j}\geqslant5$\tabularnewline
\hline 
\multirow{3}{*}{Cut-1} & Leading jet $p_{T}>200$ GeV and $|\eta|<2.0$\tabularnewline
 & Collinear jet $p_{T}>30$ GeV and $|\eta|<3.0$\tabularnewline
 & Other jets $p_{T}>30$ GeV and $|\eta|<2.5$\tabularnewline
\hline 
Cut-2 & Number of b tagged jet $N_{b}\geqslant1$\tabularnewline
\hline 
Cut-3 & Angular separation between two jets $\Delta R(j,j)>0.4$\tabularnewline
\hline 
Cut-4 & Scalar sum of $p_{T}$ of jets $H_{T}>400$ GeV\tabularnewline
\hline 
Cut-5 & Invariant mass interval of two jets $|m_{jj}-m_{W}|<20$ GeV\tabularnewline
\hline 
Cut-6 & Invariant mass interval of three jets $|m_{bjj}-m_{Y}|<0.2m_{Y}$
GeV\tabularnewline
\hline 
\end{tabular}
\end{table}

\begin{table}
\caption{Cut efficiencies (\%) for different signal benchmarks ( $\kappa_{Y}=0.5$
and $m_{Y}=1000,1500,2000,2500$ GeV) and backgrounds. \label{tab:table3}}

\begin{tabular}{|c|c|c|c|c|c|}
\hline 
 & Cut-1 & Cut-2 & Cut-3 & Cut-4 & Cut-5\tabularnewline
\hline 
\hline 
$m_{Y}=1000$ GeV & $30.28$ & $25.20$ & $25.03$ & $25.03$ & $7.60$\tabularnewline
\hline 
$m_{Y}=1500$ GeV & $30.41$ & $24.42$ & $24.35$ & $24.35$ & $4.09$\tabularnewline
\hline 
$m_{Y}=2000$ GeV & $31.84$ & $25.31$ & $25.27$ & $25.27$ & $2.58$\tabularnewline
\hline 
$m_{Y}=2500$ GeV & $33.52$ & $25.71$ & $25.71$ & $25.71$ & $2.15$\tabularnewline
\hline 
$t\bar{t}$ & $6.58$ & $4.14$ & $4.14$ & $4.14$ & $2.09$\tabularnewline
\hline 
$tb$ & $7.63$ & $6.80$ & $6.71$ & $6.71$ & $3.22$\tabularnewline
\hline 
$tW$ & $13.78$ & $6.31$ & $6.28$ & $6.28$ & $3.02$\tabularnewline
\hline 
$tj$ & $2.88$ & $1.87$ & $1.87$ & $1.87$ & $1.01$\tabularnewline
\hline 
$tZj$ & $8.37$ & $4.25$ & $4.25$ & $4.25$ & $1.72$\tabularnewline
\hline 
$Wbj$ & $3.13$ & $2.30$ & $2.30$ & $2.30$ & $0.21$\tabularnewline
\hline 
$Wbjj$ & $4.48$ & $2.65$ & $2.65$ & $2.65$ & $0.71$\tabularnewline
\hline 
$Wj$ & $1.83$ & $0.61$ & $0.61$ & $0.61$ & $0.61$\tabularnewline
\hline 
$Wjj$ & $4.53$ & $0.40$ & $0.40$ & $0.40$ & $0.10$\tabularnewline
\hline 
$Wjjj$ & $6.65$ & $0.76$ & $0.76$ & $0.76$ & $0.27$\tabularnewline
\hline 
$Zbj$ & $6.67$ & $3.74$ & $3.74$ & $3.74$ & $1.72$\tabularnewline
\hline 
$Zbjj$ & $11.22$ & $5.83$ & $5.83$ & $5.83$ & $2.49$\tabularnewline
\hline 
$Zj$ & $3.29$ & $0.00$ & $0.00$ & $0.00$ & $0.00$\tabularnewline
\hline 
$Zjj$ & $4.51$ & $0.88$ & $0.88$ & $0.88$ & $0.10$\tabularnewline
\hline 
$Zjjj$ & $6.53$ & $0.69$ & $0.69$ & $0.69$ & $0.20$\tabularnewline
\hline 
$WW$ & $7.10$ & $0.81$ & $0.81$ & $0.81$ & $0.00$\tabularnewline
\hline 
$ZZ$ & $5.03$ & $0.41$ & $0.41$ & $0.41$ & $0.27$\tabularnewline
\hline 
$WZ$ & $5.54$ & $0.66$ & $0.66$ & $0.66$ & $0.66$\tabularnewline
\hline 
\end{tabular}
\end{table}

For calculating signal significance ($SS$) as incorporated in \citep{key-17,key-18,key-19},
we use the invariant mass interval defined as $|m_{bjj}-m_{Y}|<0.2m_{Y}$.
The expected signal significance is given in terms of signal ($S$)
and background events ($B$).

In the special case of well known background, the statistical significance
($SS$)

\[
SS=\sqrt{2\left[(S+B)\ln\left(1+\frac{S}{B}\right)-S\right]}
\]
which would further reduces to $S/\sqrt{B}$ in the limit of large
background events. The signal significances for the couplings $\kappa_{Y}=0.5$,
$\kappa_{Y}=0.3$ and $\kappa_{Y}=0.15$ at the projected integrated
luminosities of $L_{int}=300$ fb$^{-1}$, $1000$ fb$^{-1}$ and
$3000$ fb$^{-1}$ are given in Table \ref{tab:table3}, \ref{tab:table4}
and \ref{tab:table5}, respectively. These results are obtained for
the invariant mass $m_{Y}$ variable after applying the cuts as in
Table \ref{tab:table2} to all signals and backgrounds event samples.

\begin{table}
\caption{Statistical significance ($SS$) for different signal benchmarks (
$\kappa_{Y}=0.5$ and $m_{Y}=1000,1500,2000,2500$ GeV) and different
integrated luminosity projections at HL-LHC. \label{tab:table4}}

\begin{tabular}{|c|c|c|c|}
\hline 
\multirow{2}{*}{Signal} & \multicolumn{3}{c|}{$L_{int}($fb$^{-1}$)}\tabularnewline
\cline{2-4} \cline{3-4} \cline{4-4} 
 & $300$ & $1000$ & $3000$\tabularnewline
\hline 
\hline 
$\kappa_{Y}=0.5$, $m_{Y}=1000\text{ GeV}$ & $9.13$ & $16.67$ & $28.88$\tabularnewline
\hline 
$\kappa_{Y}=0.5$, $m_{Y}=1500\text{ GeV}$ & $3.73$ & $6.80$ & $11.78$\tabularnewline
\hline 
$\kappa_{Y}=0.5$, $m_{Y}=2000\text{ GeV}$ & $1.33$ & $2.43$ & $4.21$\tabularnewline
\hline 
$\kappa_{Y}=0.5$, $m_{Y}=2500\text{ GeV}$ & $0.50$ & $0.92$ & $1.59$\tabularnewline
\hline 
\end{tabular}
\end{table}

\begin{table}
\noindent \caption{The same as Table \ref{tab:table4}, but for $\kappa_{Y}=0.3$ and
$m_{Y}=500,1000,1500,2000$ GeV.\label{tab:table5}}

\begin{tabular}{|c|c|c|c|}
\hline 
\multirow{2}{*}{Signal} & \multicolumn{3}{c|}{$L_{int}($fb$^{-1}$)}\tabularnewline
\cline{2-4} \cline{3-4} \cline{4-4} 
 & $300$ & $1000$ & $3000$\tabularnewline
\hline 
\hline 
$\kappa_{Y}=0.3$, $m_{Y}=500\text{ GeV}$ & $13.54$ & $24.73$ & $42.83$\tabularnewline
\hline 
$\kappa_{Y}=0.3$, $m_{Y}=1000\text{ GeV}$ & $3.22$ & $5.88$ & $10.18$\tabularnewline
\hline 
$\kappa_{Y}=0.3$, $m_{Y}=1500\text{ GeV}$ & $0.72$ & $1.32$ & $2.28$\tabularnewline
\hline 
$\kappa_{Y}=0.3$, $m_{Y}=2000\text{ GeV}$ & $0.50$ & $0.92$ & $1.59$\tabularnewline
\hline 
\end{tabular}
\end{table}

\begin{table}
\caption{The same as Table \ref{tab:table4}, but for $\kappa_{Y}=0.15$ and
$m_{Y}=500,1000,1500,2000$ GeV.\label{tab:table6}}

\begin{tabular}{|c|c|c|c|}
\hline 
\multirow{2}{*}{Signal} & \multicolumn{3}{c|}{$L_{int}($fb$^{-1}$)}\tabularnewline
\cline{2-4} \cline{3-4} \cline{4-4} 
 & $300$ & $1000$ & $3000$\tabularnewline
\hline 
\hline 
$\kappa_{Y}=0.15$, $m_{Y}=500\text{ GeV}$ & $3.60$ & $6.58$ & $11.39$\tabularnewline
\hline 
$\kappa_{Y}=0.15$, $m_{Y}=1000\text{ GeV}$ & $0.77$ & $1.41$ & $2.45$\tabularnewline
\hline 
$\kappa_{Y}=0.15$, $m_{Y}=1500\text{ GeV}$ & $0.17$ & $0.31$ & $0.53$\tabularnewline
\hline 
$\kappa_{Y}=0.15$, $m_{Y}=2000\text{ GeV}$ & $0.15$ & $0.28$ & $0.48$\tabularnewline
\hline 
\end{tabular}
\end{table}

\begin{figure}
\includegraphics[scale=0.8]{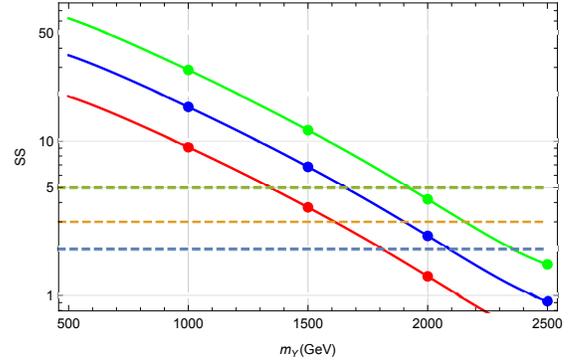}

\caption{The statistical significance $SS$ for $\kappa_{Y}=0.5$ as a function of vectorlike $Y$
quark mass $m_{Y}$ for integrated luminosity projections (solid lines,
red to green) $L_{int}=300$ fb$^{-1}$, $1000$ fb$^{-1}$ and $3000$
fb$^{-1}$ at HL-LHC. The horizontal dashed lines correspond to $2\sigma$,
$3\sigma$ and $5\sigma$ significances (lower to upper), respectively.
\label{fig:fig10}}
\end{figure}

\begin{figure}
\includegraphics[scale=0.8]{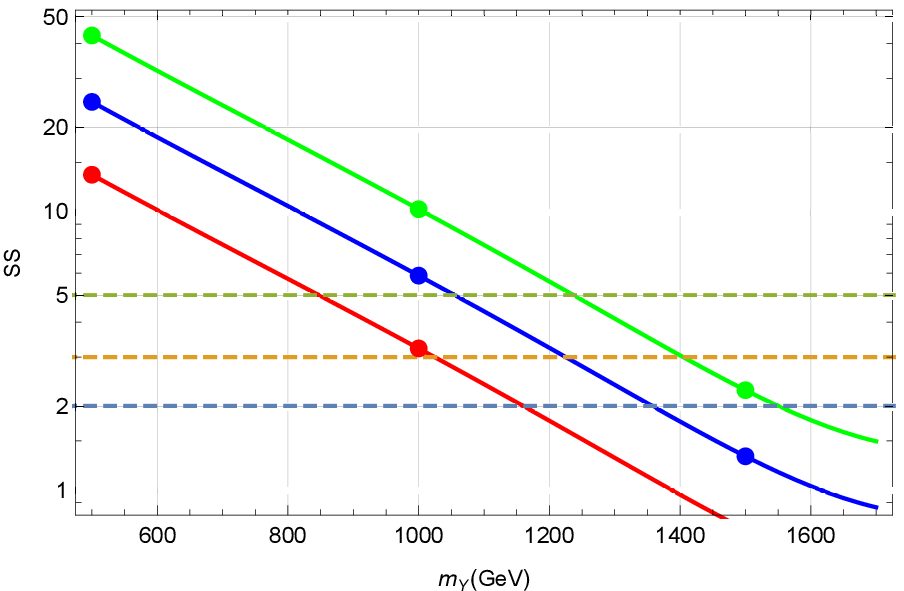}

\caption{The same as Fig. \ref{fig:fig10}, but for $\kappa_{Y}=0.3$.\label{fig:fig11}}
\end{figure}

\begin{figure}
\includegraphics[scale=0.8]{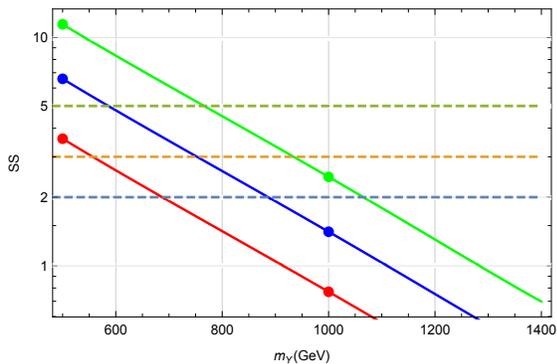}

\caption{The same as Fig. \ref{fig:fig10}, but for $\kappa_{Y}=0.15$.\label{fig:fig12}}
\end{figure}

The statistical significance $SS$ as a function of mass of vectorlike
$Y$ quark for $L_{int}=300$ fb$^{-1}$, $1000$ fb$^{-1}$ and $3000$
fb$^{-1}$ are shown in Fig. \ref{fig:fig10}, \ref{fig:fig11} and
\ref{fig:fig12}. In these figures, the coupling values $\kappa_{Y}=0.5$,
$\kappa_{Y}=0.3$ and $\kappa_{Y}=0.15$ are chosen as benchmarks,
respectively. For the integrated luminosity projection of $3000$
fb$^{-1}$, lower limits for $m_{Y}$ are obtained as $2350$ GeV
at $2\sigma$ significance level, $2150$ GeV at $3\sigma$ (observability)
and $1900$ GeV at $5\sigma$ (discovery) significances. For other
couplings $\kappa_{Y}=0.3$ and $\kappa_{Y}=0.15$ the lower limits
for $m_{Y}$ are found as $1550$ GeV and $1075$ GeV at $2\sigma$
significance level at $L_{int}=3000$ fb$^{-1}$, respectively.

\section{Conclusions}
We have studied single production of vectorlike $Y$ quark and its
subsequent decay ($Y\to Wb$) with phenomenological interpretations
in the context of well motivated VLQ model framework. Signal significances
have been obtained depending on the parameter space (mass in the range
$1000-2500$ GeV for coupling $\kappa_{Y}=0.5$ and $500-2000$ GeV
for couplings $\kappa_{Y}=0.3$ and $\kappa_{Y}=0.15$) using all
hadronic mode (at least five jets final state). We find a significant
coverage of the signal parameter space and distinguish the signal
for a mass up to $2350$ GeV for an integrated luminosity of $3000$
fb$^{-1}$ at the HL-LHC. We have used the criteria for projected
discovery \citep{key-20} sensitivities to the couplings and masses.
For signal significance $2\sigma$ level we find vectorlike $Y$ quark
attainable mass limits as $1800$ GeV, $2100$ GeV and $2350$ GeV
at projected integrated luminosities $L_{int}=300$ fb$^{-1},$$1000$
fb$^{-1}$ and $3000$ fb$^{-1}$, respectively. For other coupling
values $\kappa_{Y}=0.3$ ($\kappa_{Y}=0.15$) we find attainable mass
limits as $1150$ GeV ($700$ GeV), $1350$ GeV ($900$ GeV) and $1550$
GeV ($1075$ GeV) at the same integrated luminosity projections.
The systematics are known to affect the discovery reaches.
However, given the difficulty to perform a precise estimation 
of systematics for projected results, the significances have been 
computed without considering the systematic uncertainties (an optimistic case) 
for the estimation of the potential of HL-LHC.
These results can also be used for constraining the models 
\citep{key-21,key-22} predicting new heavy quarks with exotic charge as $-4/3$.
This analysis shows that HL-LHC could discover a wide range of parameter
space of vectorlike $Y$ quark models.

\begin{acknowledgments}
The numerical calculations reported in this paper were partially performed
at TUBITAK ULAKBIM, High Performance and Grid Computing Center (TRUBA
resources). The work of O.C. was supported in part by the Turkish
Atomic Energy Authority (TAEA) under grant No. 2020TAEK(CERN)A5.H1.F5-25. 
\end{acknowledgments}

\appendix
\section{}

We present following Table \ref{tab:tableA1} and Figures
\ref{fig:figA1} - \ref{fig:figA9} to avoid detailed information in the main text.  
 
\setcounter{table}{0}
\renewcommand\thetable{\Alph{section}.\arabic{table}}

\setcounter{figure}{0}
\renewcommand\thefigure{\Alph{section}.\arabic{figure}}

 \begin{table}[b]
\caption{The leading order (LO) signal cross section values (in pb) for the
process $pp\rightarrow Ybj+X$ at the HL-LHC with $\sqrt{s}=14$ TeV.
\label{tab:tableA1}}

\begin{ruledtabular}
\tiny%
\begin{tabular}{|c|c|c|c|c|}
\hline 
$m_{Y}(\text{GeV})$ & $\kappa_{Y}=0.5$ & $\kappa_{Y}=0.2$ & $\kappa_{Y}=0.1$ & $\kappa_{Y}=0.05$\tabularnewline
\hline 
\hline 
$1000$ & $1.08\times10^{0}$ & $1.72\times10^{-1}$ & $4.30\times10^{-2}$ & $1.08\times10^{-2}$\tabularnewline
\hline 
$1100$ & $8.01\times10^{-1}$ & $1.28\times10^{-1}$ & $3.21\times10^{-2}$ & $8.01\times10^{-3}$\tabularnewline
\hline 
$1200$ & $6.04\times10^{-1}$ & $9.63\times10^{-2}$ & $2.42\times10^{-2}$ & $6.04\times10^{-3}$\tabularnewline
\hline 
$1300$ & $4.61\times10^{-1}$ & $7.37\times10^{-2}$ & $1.85\times10^{-2}$ & $4.60\times10^{-3}$\tabularnewline
\hline 
$1400$ & $3.55\times10^{-1}$ & $5.68\times10^{-2}$ & $1.41\times10^{-2}$ & $3.55\times10^{-3}$\tabularnewline
\hline 
$1500$ & $2.76\times10^{-1}$ & $4.41\times10^{-2}$ & $1.10\times10^{-2}$ & $2.75\times10^{-3}$\tabularnewline
\hline 
$1600$ & $2.15\times10^{-1}$ & $3.44\times10^{-2}$ & $8.56\times10^{-3}$ & $2.14\times10^{-3}$\tabularnewline
\hline 
$1700$ & $1.69\times10^{-1}$ & $2.70\times10^{-2}$ & $6.73\times10^{-3}$ & $1.69\times10^{-3}$\tabularnewline
\hline 
$1800$ & $1.33\times10^{-1}$ & $2.13\times10^{-2}$ & $5.30\times10^{-3}$ & $1.33\times10^{-3}$\tabularnewline
\hline 
$1900$ & $1.06\times10^{-1}$ & $1.69\times10^{-2}$ & $4.21\times10^{-3}$ & $1.05\times10^{-3}$\tabularnewline
\hline 
$2000$ & $8.44\times10^{-2}$ & $1.34\times10^{-2}$ & $3.37\times10^{-3}$ & $8.41\times10^{-4}$\tabularnewline
\hline 
$2100$ & $6.72\times10^{-2}$ & $1.07\times10^{-2}$ & $2.70\times10^{-3}$ & $6.71\times10^{-4}$\tabularnewline
\hline 
$2200$ & $5.39\times10^{-2}$ & $8.62\times10^{-3}$ & $2.16\times10^{-3}$ & $5.40\times10^{-4}$\tabularnewline
\hline 
$2300$ & $4.34\times10^{-2}$ & $6.95\times10^{-3}$ & $1.74\times10^{-3}$ & $4.34\times10^{-4}$\tabularnewline
\hline 
$2400$ & $3.49\times10^{-2}$ & $5.62\times10^{-3}$ & $1.40\times10^{-3}$ & $3.50\times10^{-4}$\tabularnewline
\hline 
$2500$ & $2.83\times10^{-2}$ & $4.54\times10^{-3}$ & $1.13\times10^{-3}$ & $2.84\times10^{-4}$\tabularnewline
\hline 
$2600$ & $2.29\times10^{-2}$ & $3.67\times10^{-3}$ & $9.11\times10^{-4}$ & $2.29\times10^{-4}$\tabularnewline
\hline 
$2700$ & $1.86\times10^{-2}$ & $2.98\times10^{-3}$ & $7.42\times10^{-4}$ & $1.85\times10^{-4}$\tabularnewline
\hline 
$2800$ & $1.51\times10^{-2}$ & $2.42\times10^{-3}$ & $6.02\times10^{-4}$ & $1.51\times10^{-4}$\tabularnewline
\hline 
$2900$ & $1.23\times10^{-2}$ & $1.97\times10^{-3}$ & $4.91\times10^{-4}$ & $1.23\times10^{-4}$\tabularnewline
\hline 
$3000$ & $1.01\times10^{-2}$ & $1.60\times10^{-3}$ & $4.00\times10^{-4}$ & $9.98\times10^{-5}$\tabularnewline
\hline 
\end{tabular}
\end{ruledtabular}

\end{table}

\begin{figure}
\includegraphics[scale=0.4]{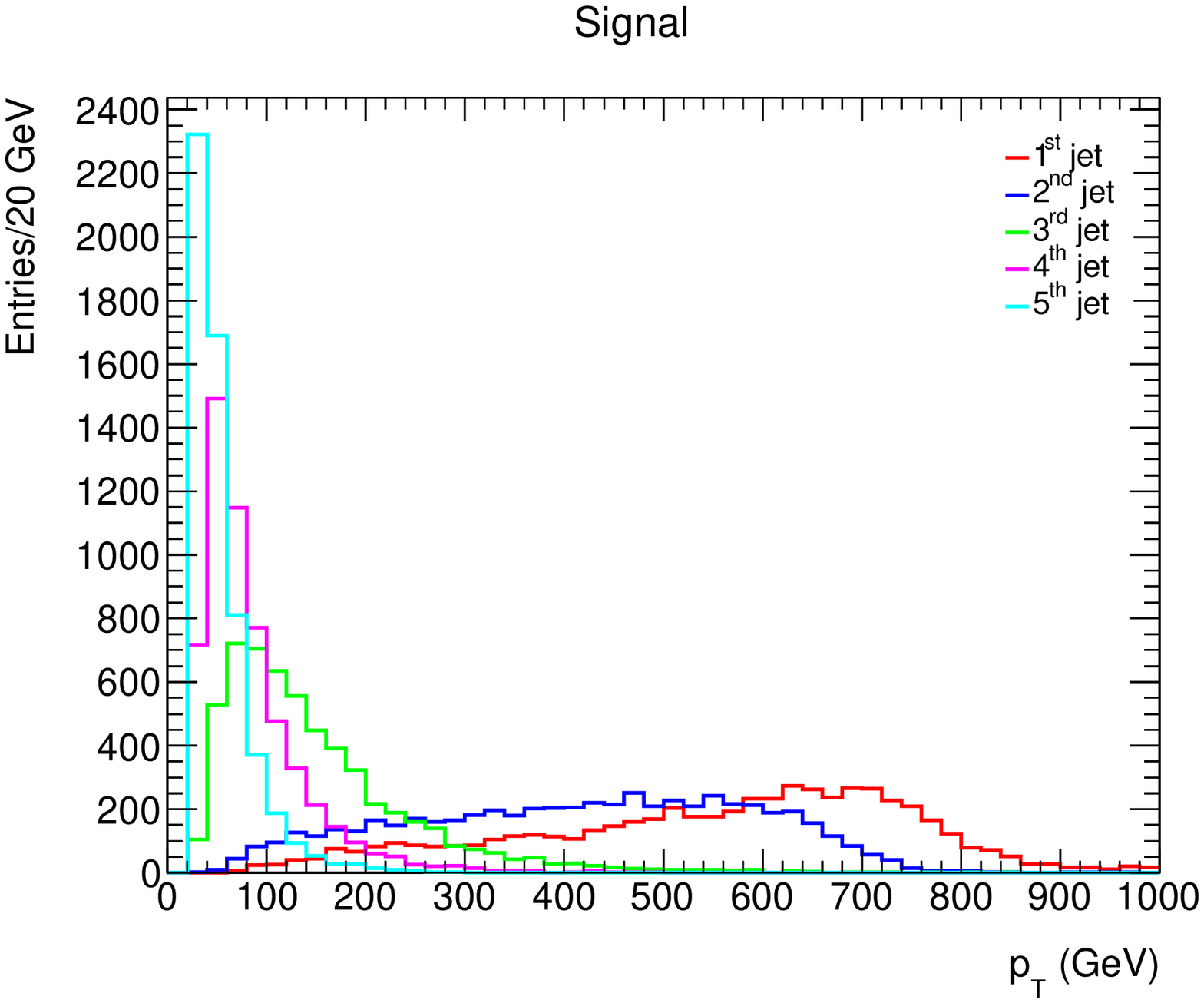}

\includegraphics[scale=0.4]{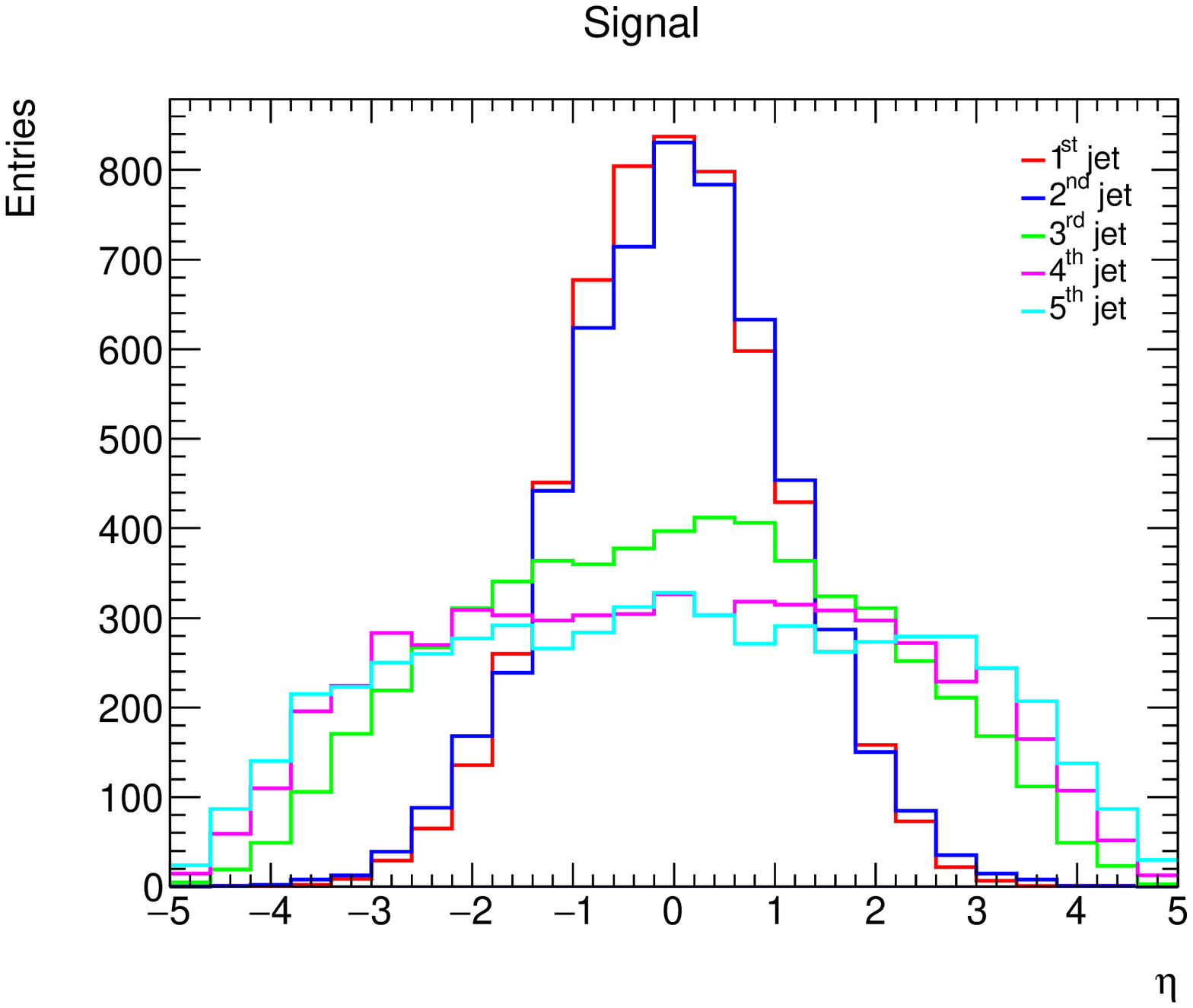}

\caption{Transverse momentum (upper) and pseudo-rapidity (lower) distribution
of five jets for signal with $m_{Y}=1500$ GeV and $\kappa_{Y}=0.5$.
\label{fig:figA1}}
\end{figure}

\begin{figure}
\includegraphics[scale=0.4]{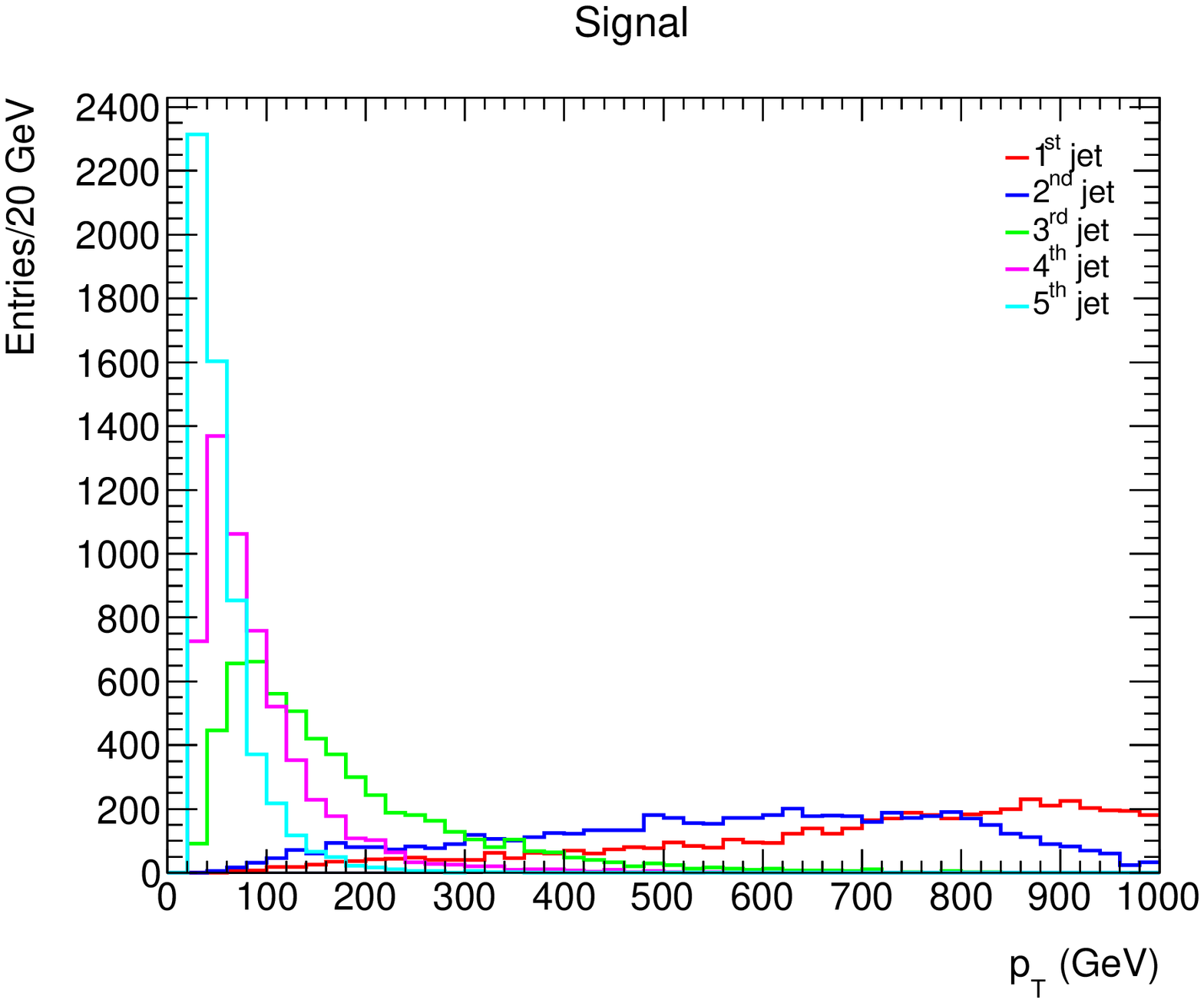}

\includegraphics[scale=0.4]{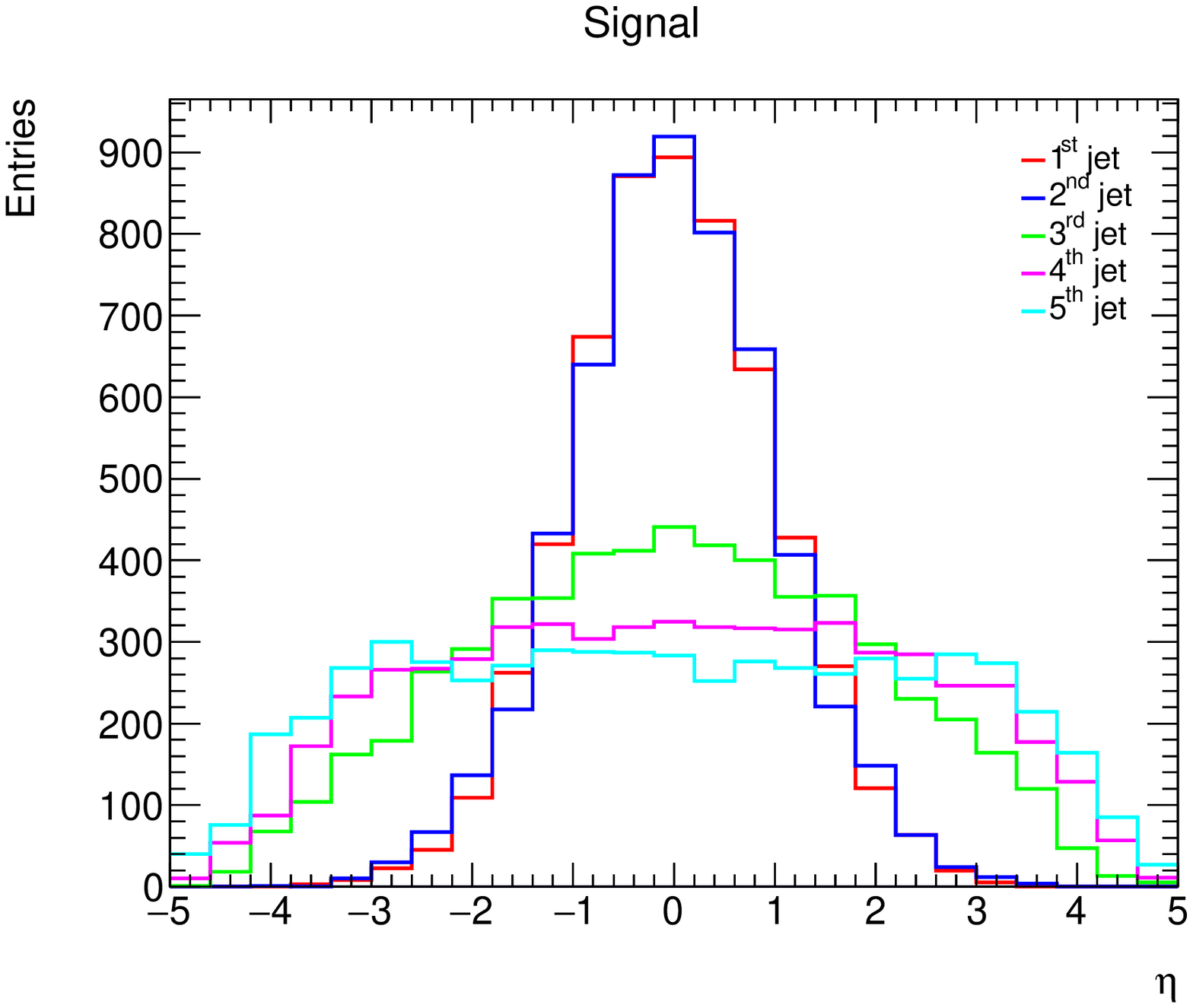}

\caption{The same as Fig. \ref{fig:figA1}, but for signal with $m_{Y}=2000$
GeV and $\kappa_{Y}=0.5$. \label{fig:figA2}}
\end{figure}

\begin{figure}[H]
\begin{raggedright}
\includegraphics[scale=0.4]{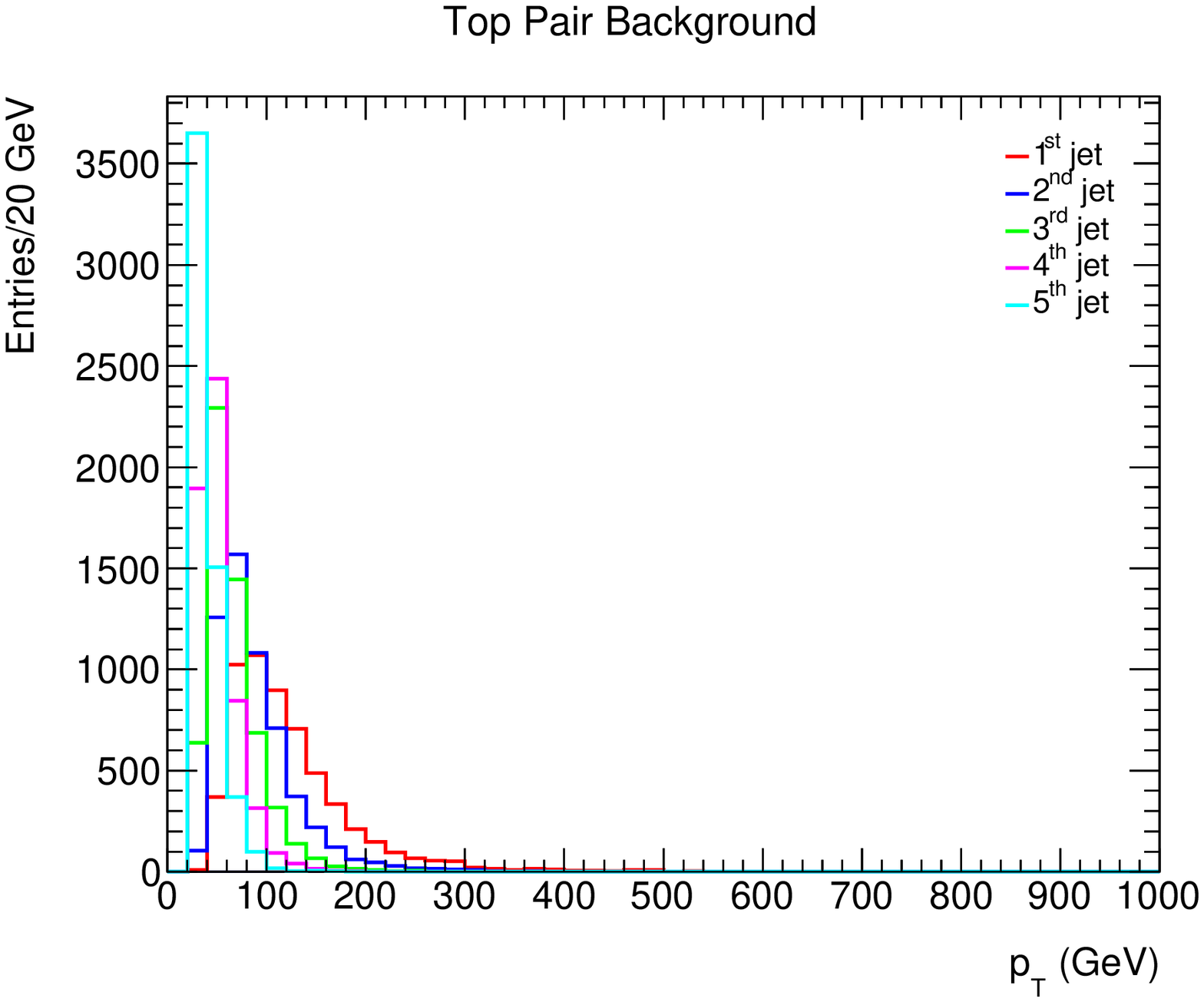}
\par\end{raggedright}
\begin{raggedright}
\includegraphics[scale=0.4]{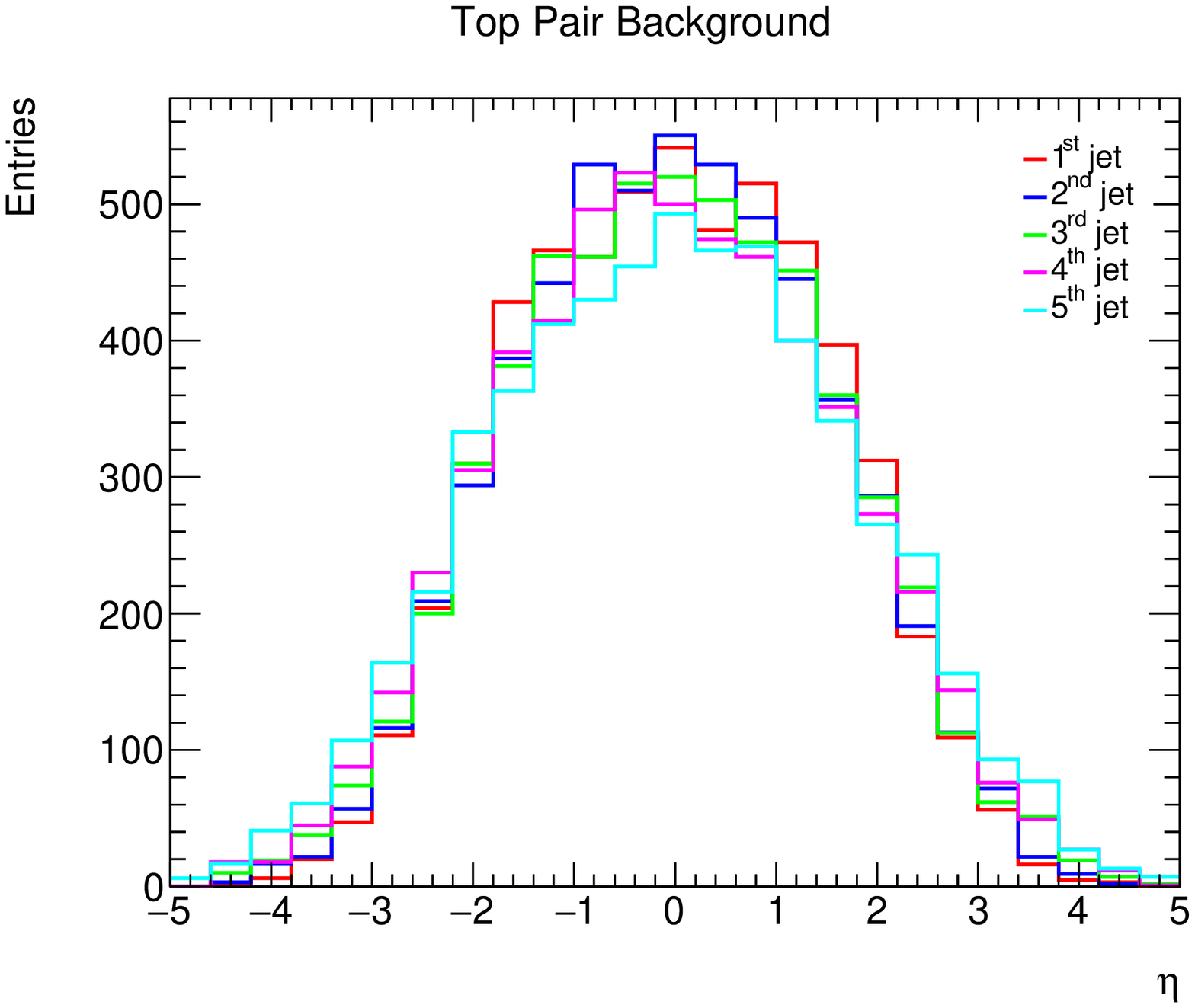}
\par\end{raggedright}
\raggedright{}\caption{Transverse momentum (upper) and pseudo-rapidity (lower) distributions
of five jets for top-pair background. \label{fig:figA3}}
\end{figure}

\begin{figure}[H]
\begin{raggedright}
\includegraphics[scale=0.4]{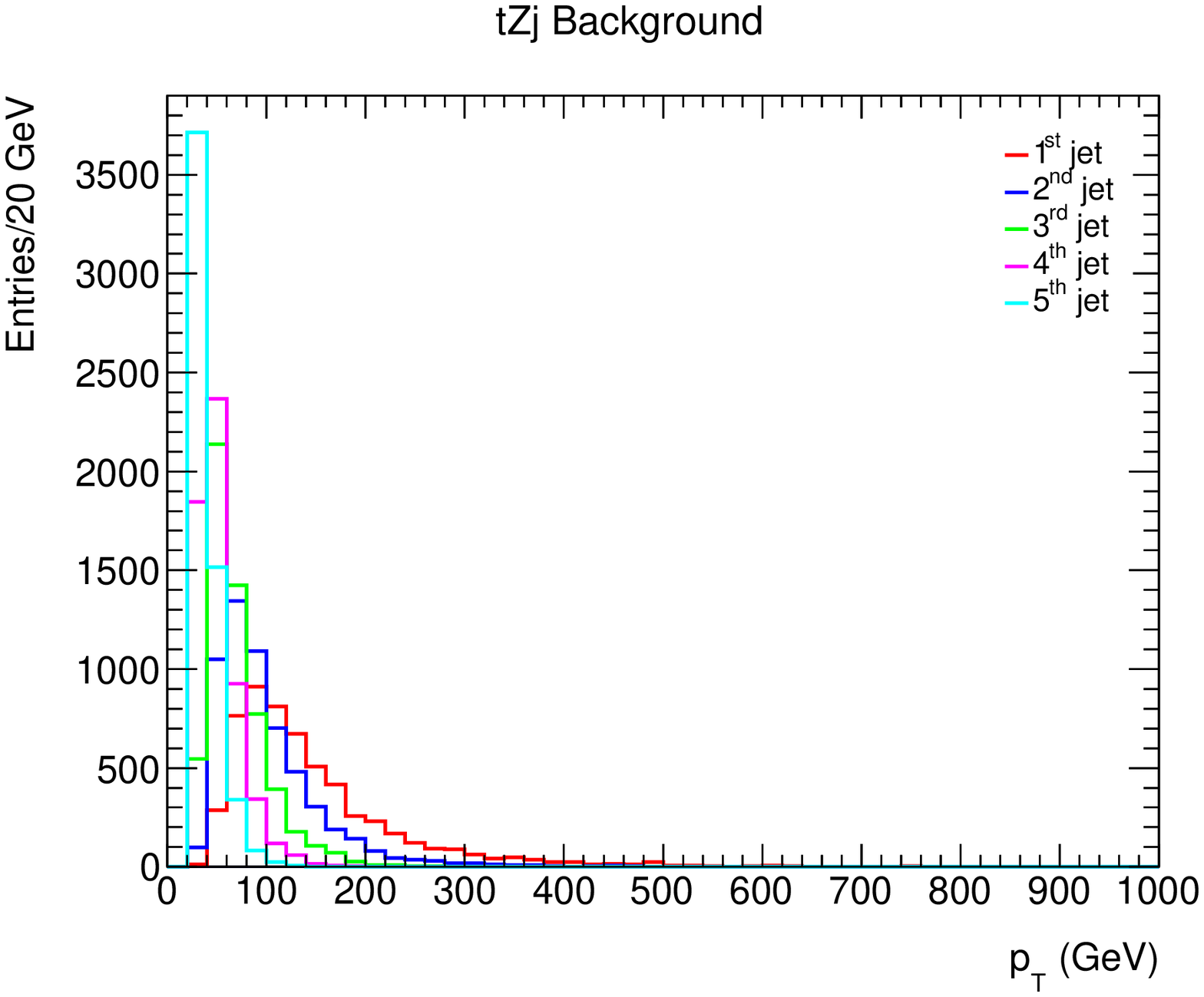}
\par\end{raggedright}
\includegraphics[scale=0.4]{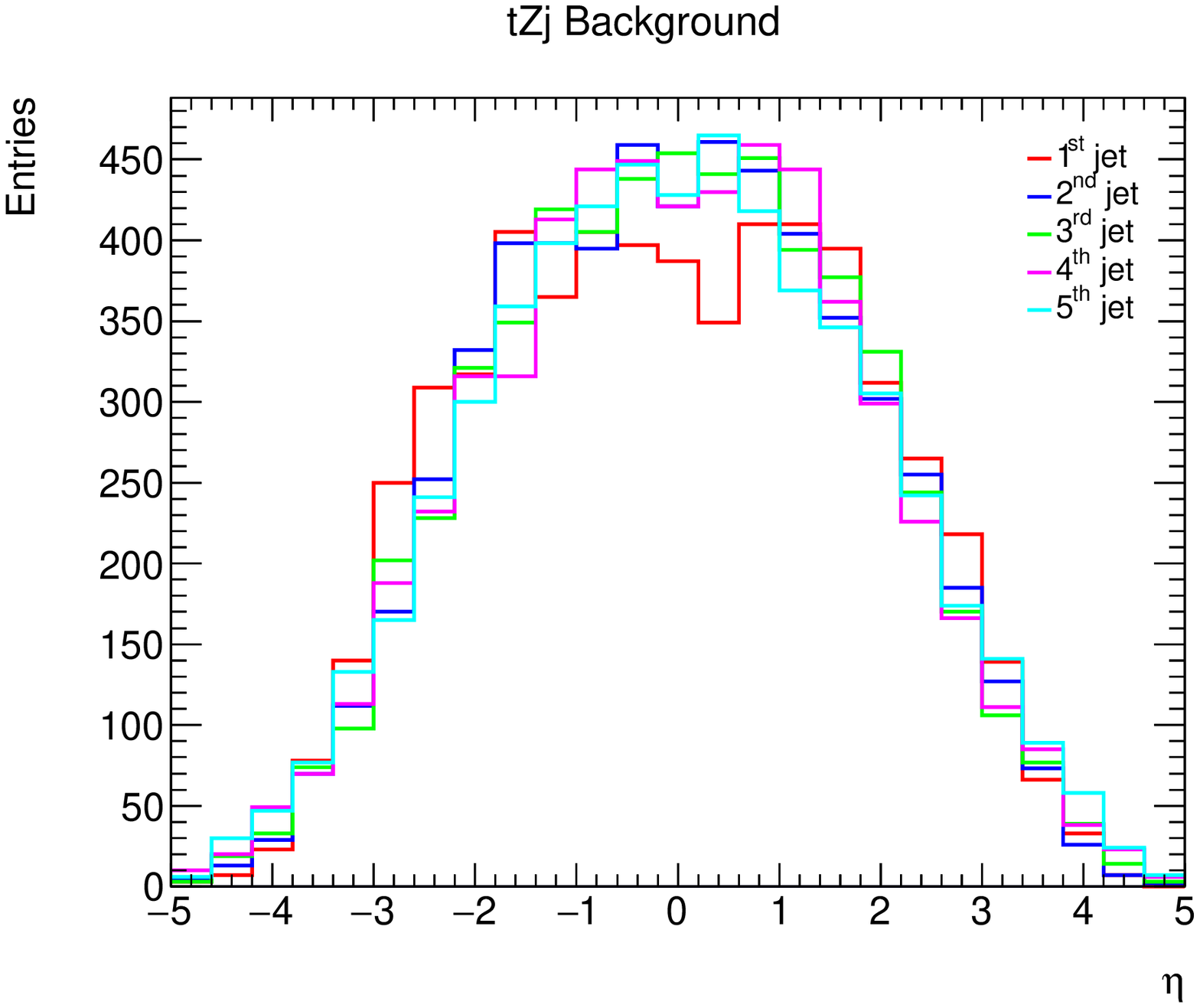}
\raggedright{}\caption{Transverse momentum (upper) and pseudo-rapidity (lower) distributions
of five jets for associated tZj production background. \label{fig:figA4}}
\end{figure}

\begin{figure}
\includegraphics[scale=0.4]{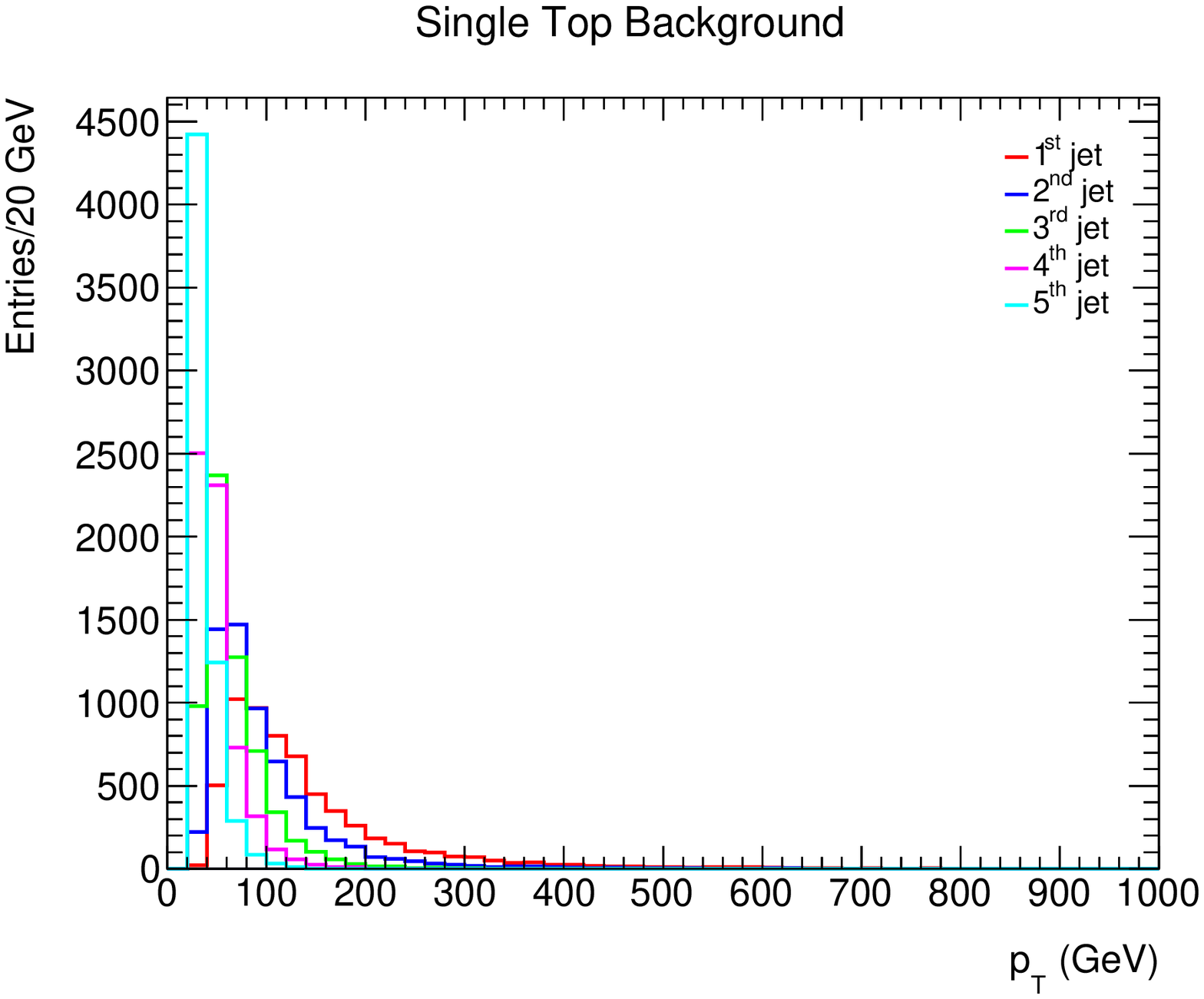}

\includegraphics[scale=0.4]{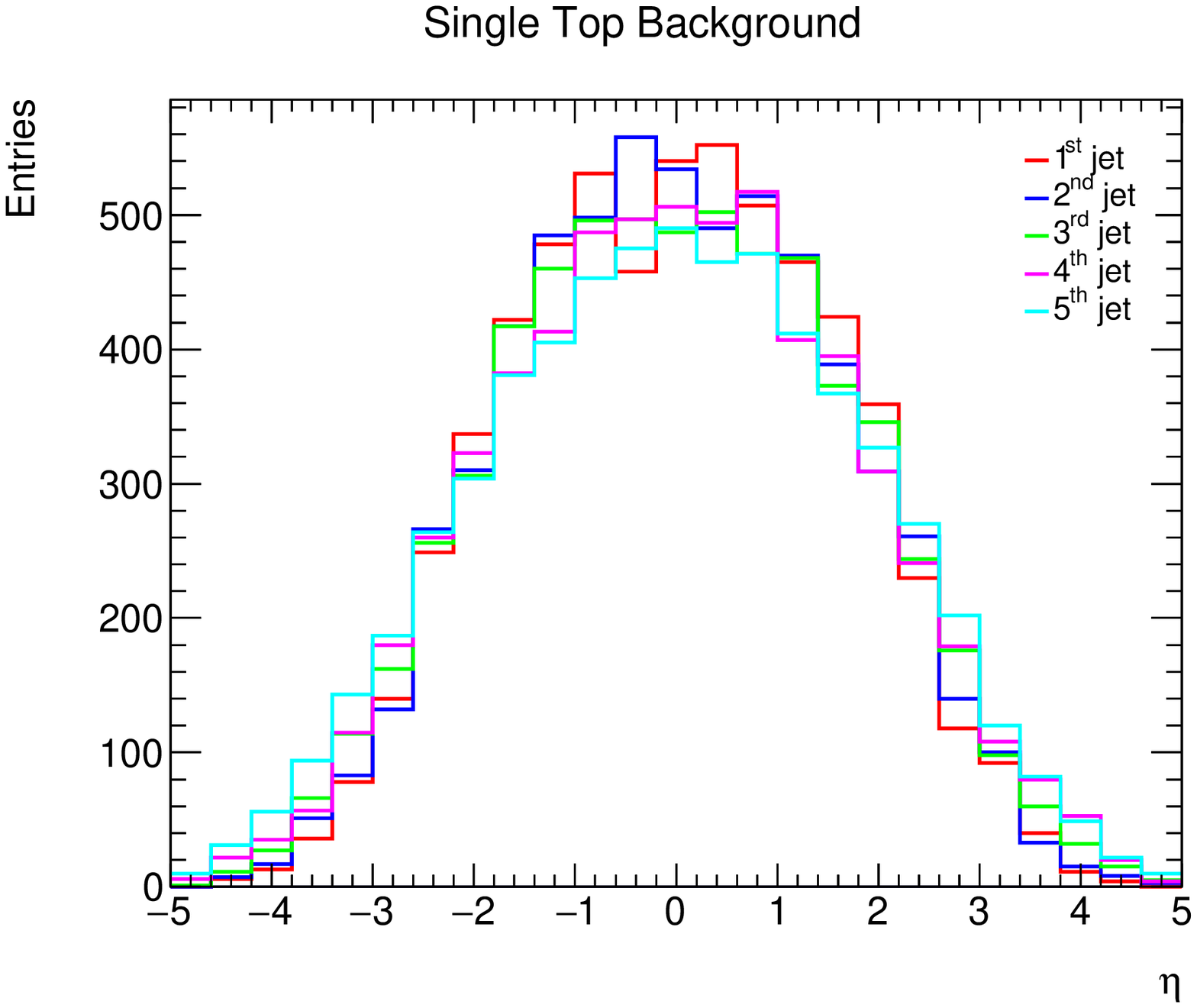}

\caption{Transverse momentum (upper) and pseudo-rapidity (lower) distributions
of five jets for single top background. \label{fig:figA5}}
\end{figure}

\begin{figure}
\includegraphics[scale=0.4]{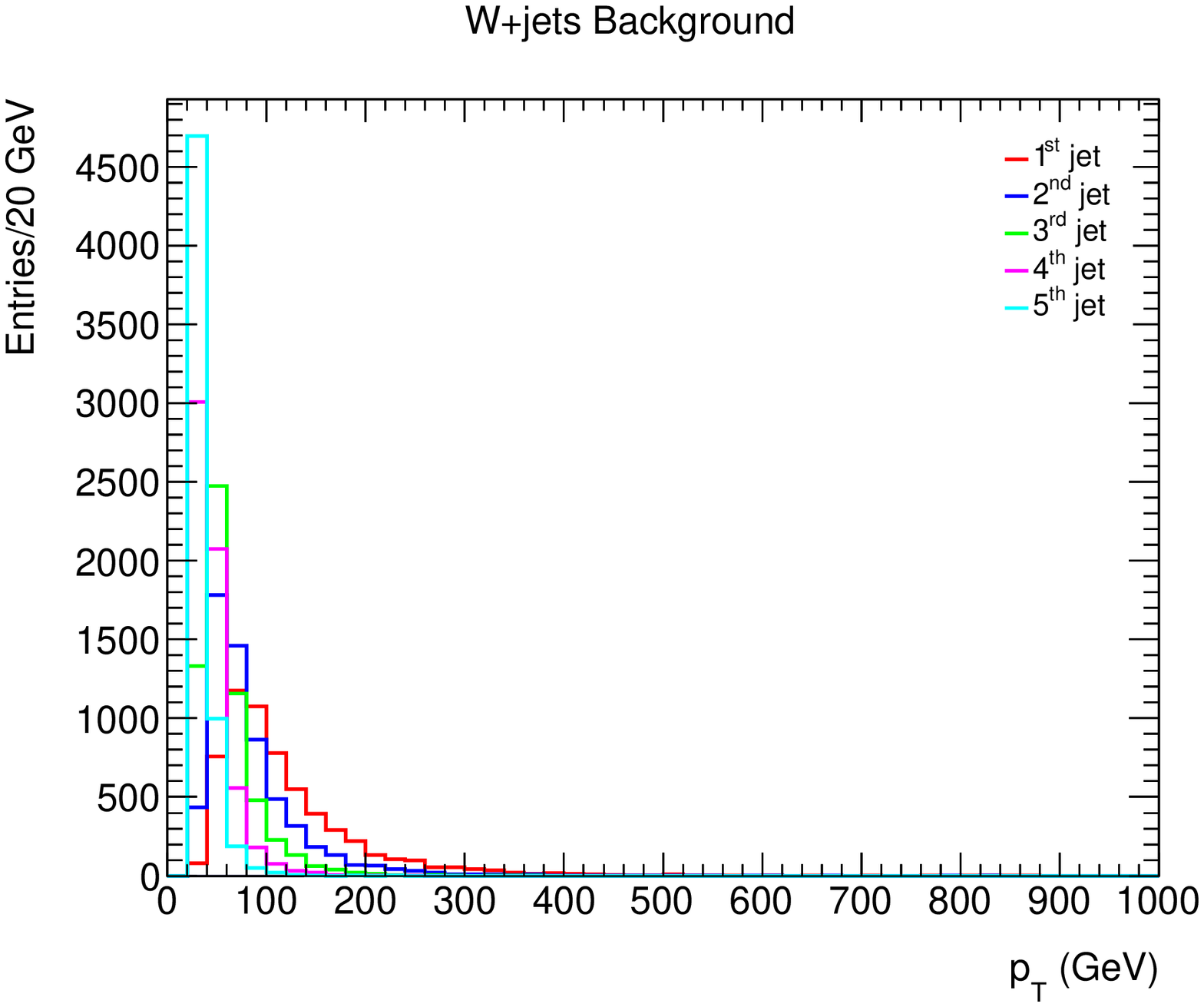}

\includegraphics[scale=0.4]{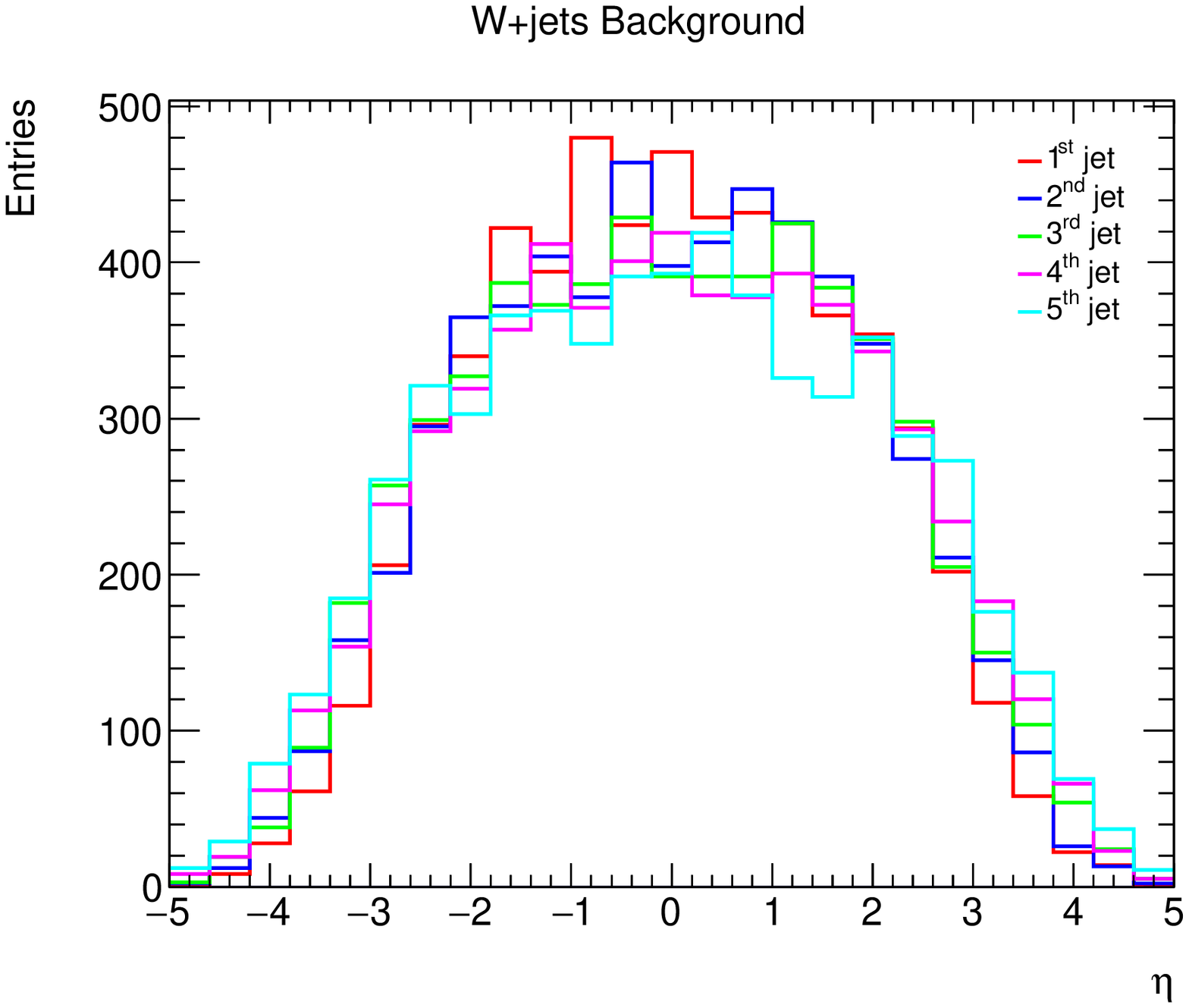}

\caption{Transverse momentum (upper) and pseudo-rapidity (lower) distributions
of five jets for $W+jets$ background. \label{fig:figA6}}
\end{figure}

\begin{figure}
\includegraphics[scale=0.4]{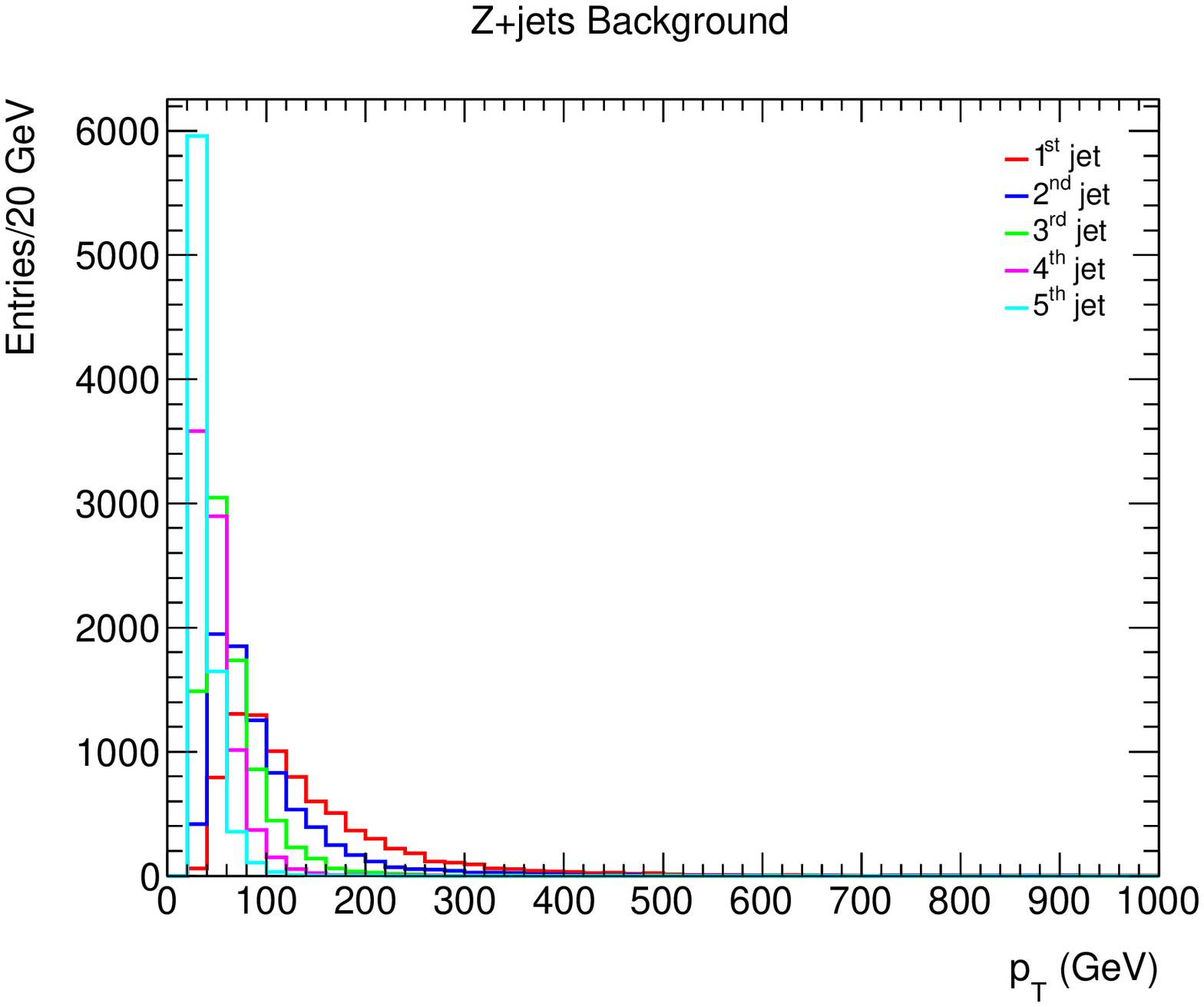}

\includegraphics[scale=0.4]{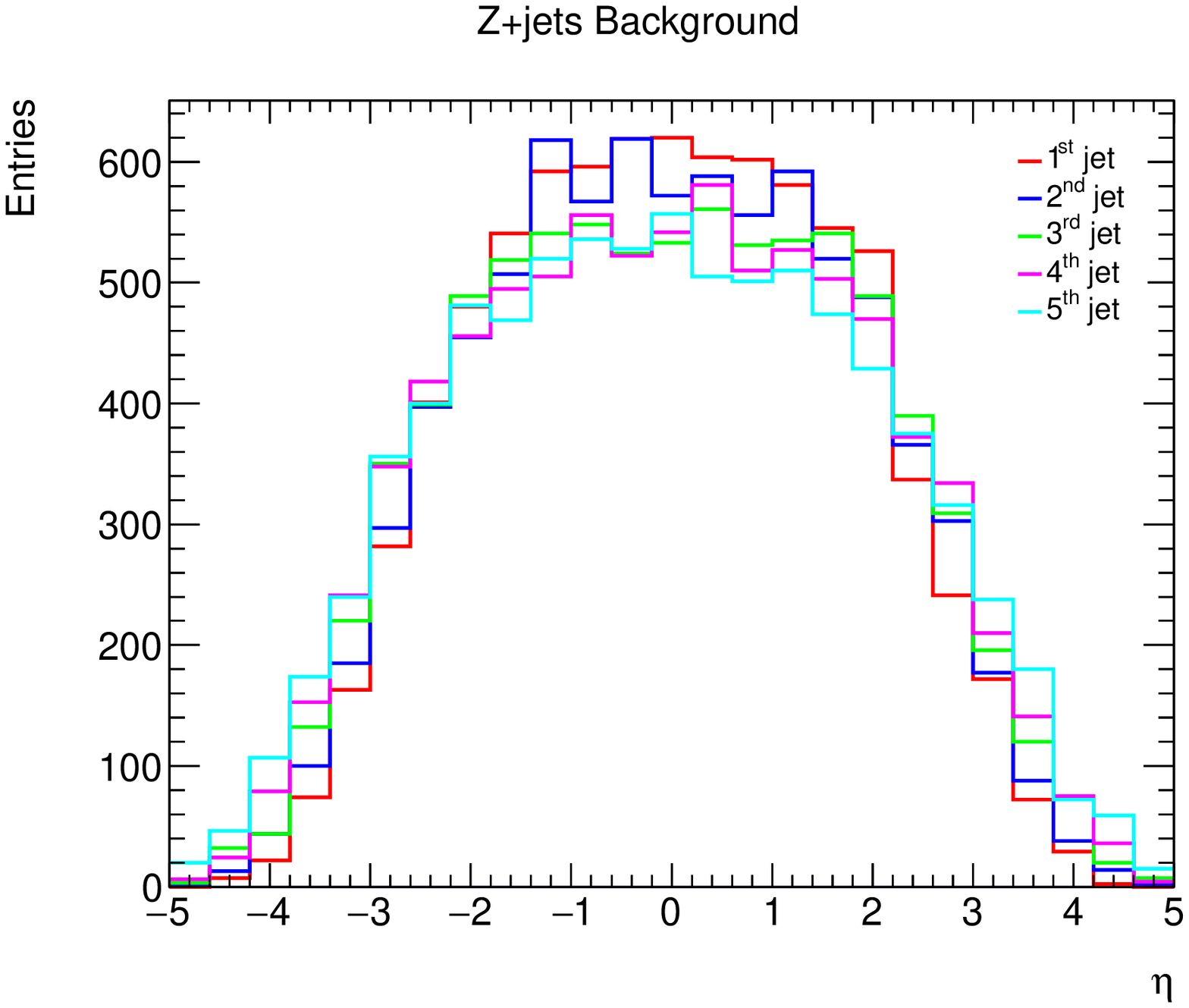}

\caption{Transverse momentum (upper) and pseudo-rapidity (lower) distributions
of five jets for $Z+jets$ background. \label{fig:figA7}}
\end{figure}

\begin{figure}
\includegraphics[scale=0.4]{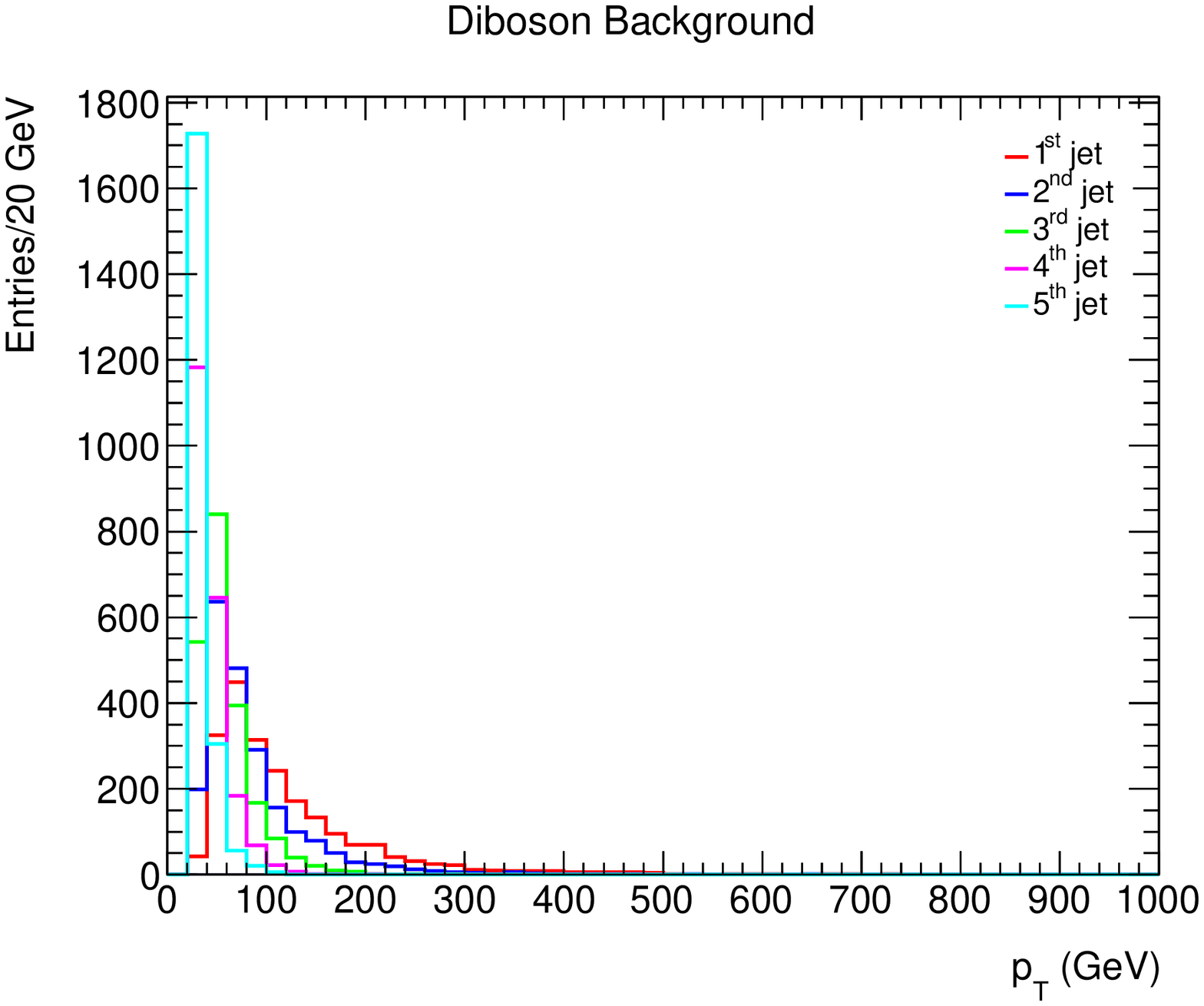}

\includegraphics[scale=0.4]{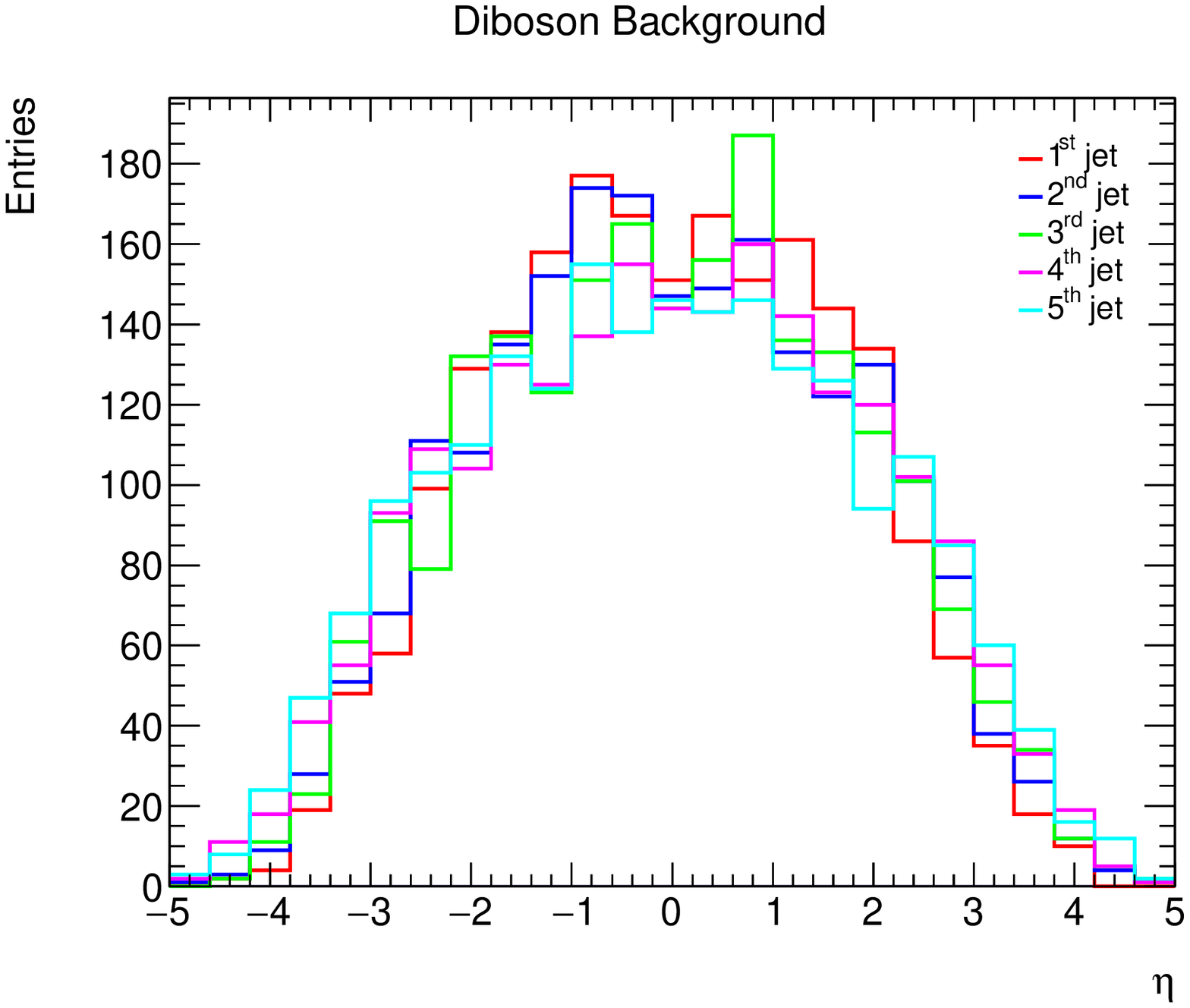}

\caption{Transverse momentum (upper) and pseudo-rapidity (lower) distributions
of five jets for diboson background. \label{fig:figA8}}
\end{figure}

\begin{figure}
\includegraphics[scale=0.4]{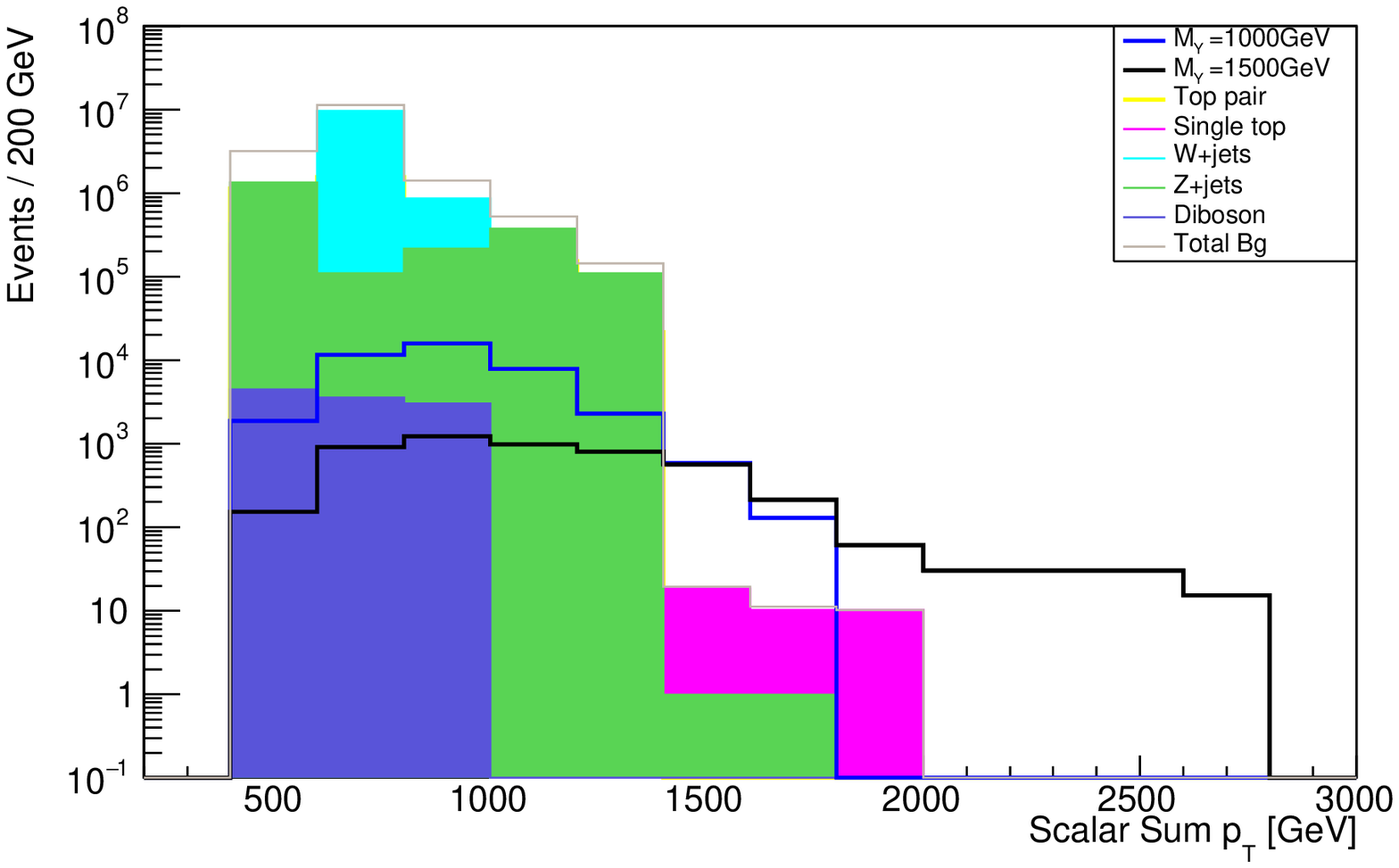}

\includegraphics[scale=0.4]{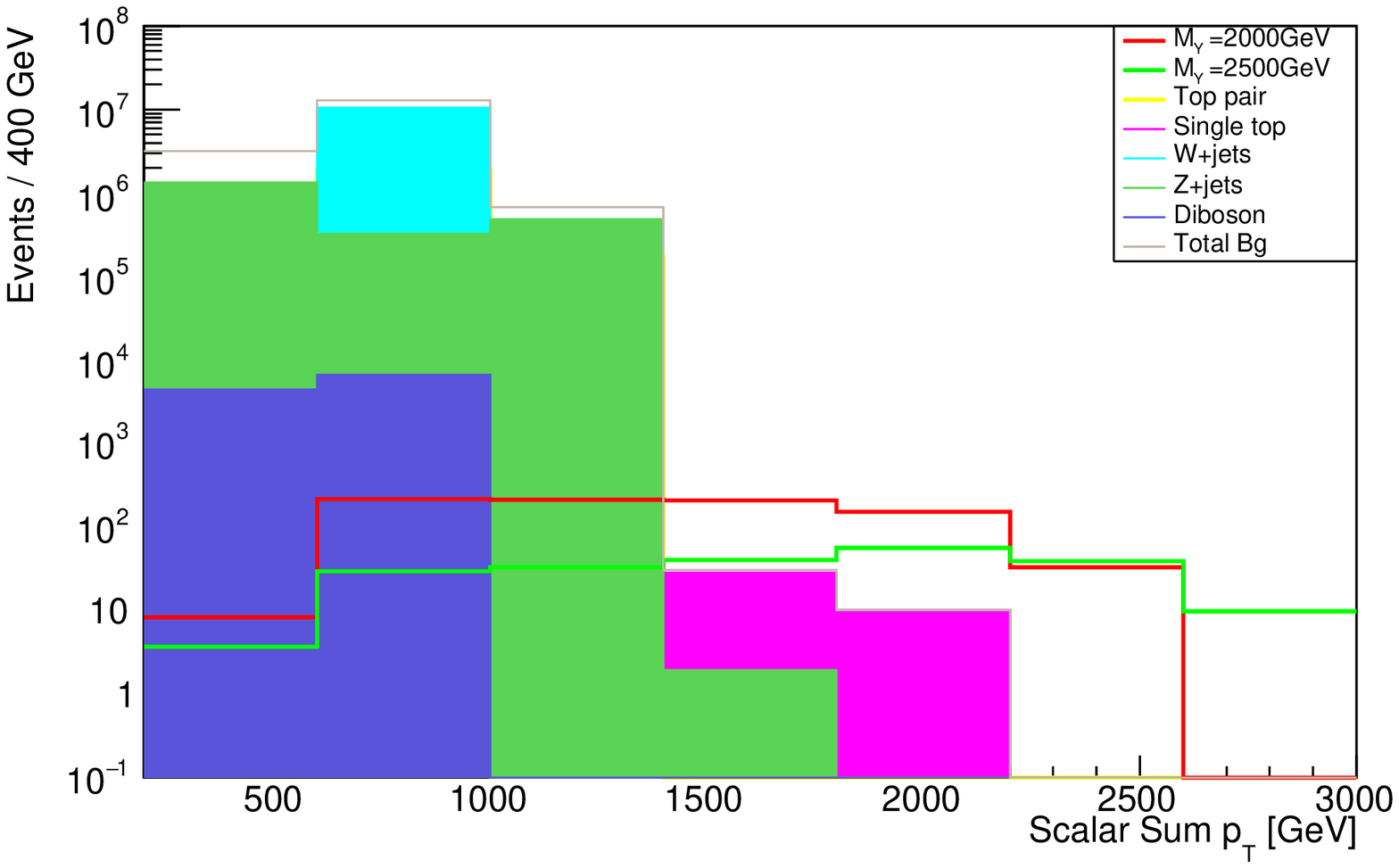}

\caption{The scalar sum $p_{T}$ for all signals ($\kappa_{Y}=0.5$) and backgrounds.
For $m_{Y}=1000$ and $1500$ GeV (upper), for $m_{Y}=2000$ and $2500$
GeV (lower). \label{fig:figA9}}
\end{figure}


\begin{thebibliography}{99}
\bibitem{key-1}Apollinari G et al. 2017 High-Luminosity Large Hadron
Collider (HL-LHC) Technical Design Report V. 0.1 (Geneva: CERN). https://doi.org/10.23731/CYRM-2017-004.

\bibitem{key-2}M. Buchkremer, G. Cacciapaglia, A. Deandrea, L. Panizzi,
Model Independent Framework for Searches of Top Partners, hep-ph/1305.4172.

\bibitem{key-3}J. A. Aguilar-Saavedra, R. Benbrik, S. Heinemeyer
and M. Perez-Victoria, Handbook of vectorlike quarks: Mixing and single
production, Phys. Rev. D 88 (2013) 094010, arXiv: 1306.0572 {[}hep-ph{]}.

\bibitem{key-4}ATLAS Collaboration, Search for single production
of vectorlike quarks decaying into Wb in pp collisions at $\sqrt{s}=13$
TeV with the ATLAS detector, JHEP05, 164 (2019).

\bibitem{key-5}CMS Collaboration, Search for single production of
vectorlike quarks decaying into a b quark and a W boson in proton\textendash proton
collisions at $\sqrt{s}=13$ TeV, Physics Letters B 772, 634-656 (2017).

\bibitem{key-6}Alexandra Carvalho, Stefano Moretti, Dermot O\textquoteright Brien,
Luca Panizzi, Hugo Prager, Single production of vector like quarks
with large width at the Large Hadron Collider, Phys.Rev. D 98, 015029
(2018).

\bibitem{key-7}Aldo Deandrea, Thomas Flacke, Benjamin Fuks, Luca Panizzi
and Hua-Sheng Shaoc, Single production of vector-like quarks: the effects of
large width, interference and NLO corrections, JHEP08(2021)107.

\bibitem{key-8}J. A. Aguilar-Saavedra, Mixing with vectorlike quarks:
constraints and expectations, EPJ Web Conf. 60 (2013) 16012, arXiv:
1306.4432 {[}hep-ph{]}.

\bibitem{key-9}Benjamin Fuks, Hua-Sheng Shao, QCD next-to-leading-order
predictions matched to parton showers for vectorlike quark models,
Eur. Phys. J. C (2017) 77:135. arXiv: 1610.04622 {[}hep-ph{]}

\bibitem{key-10}J. Alwall, R. Frederix, S. Frixione, V. Hirschi,
F. Maltoni, O. Mattelaer, H.-S. Shao, T. Stelzer, P. Torrielli, M.
Zaro, The automated computation of tree-level and next-to-leading
order differential cross sections, and their matching to parton shower
simulations, arXiv:1405.0301 {[}hep-ph{]}.

\bibitem{key-11}A. Alloul, N.D. Christensen, C. Degrande, C. Duhr,
B. Fuks, FeynRules 2.0 \textemdash{} A complete toolbox for tree-level
phenomenology, Computer Physics Communications Volume 185, Issue 8,
Pages: 2250-2300, (2014).

\bibitem{key-12}C. Degrande, C. Duhr, B. Fuks, D. Grellscheid, O.
Mattelaer, T. Reiter, UFO \textendash{} The Universal FeynRules Output,
Computer Physics Communications, Volume 183, Issue 6, Pages: 1201-1214,
(2012).

\bibitem{key-13}Torbjörn Sjöstrand, Stefan Ask, Jesper R. Christiansen,
Richard Corke, Nishita Desai, Philip Ilten, Stephen Mrenna, Stefan
Prestel, Christine O. Rasmussen, Peter Z. Skand, An introduction to
PYTHIA 8.2, Computer Physics Communications, Volume 191, Pages 159-177
(2015).

\bibitem{key-14}R.D. Ball et al., Parton distributions with LHC data,
Nucl. Phys. B 867 (2013) 244 {[}arXiv:1207.1303{]} {[}Inspire{]}.

\bibitem{key-15}J. de Favereau, C. Delaere, P. Demin, A. Giammanco,
V. Lematre, A. Mertens and M. Selvaggi, ``Delphes 3, A modular framework
for fast simulation of a generic collider experiment'', arXiv:1307.6346
{[}hep-ex{]}.

\bibitem{key-16}R. Brun, F. Rademakers, Nuclear instruments and methods
in physics research section A: accelerators, spectrometers, detectors
and associated equipment. New Comput. Techn. Phys. Res. V 389, 81
(1997). https://doi.org/10.1016/S0168-9002(97)00048-X.

\bibitem{key-17}G. Cowan, K. Cranmer, E. Gross and O. Vitells, Asymptotic
formulae for likelihood-based tests of new physics\textquotedblright ,
Eur. Phys. J. C 71, 1554 (2011) {[}Eur. Phys. J. C 73, 2501 (2013){]}
{[}arXiv:1007.1727 {[}physics.data-an{]}{]}.

\bibitem{key-18}Robert D. Cousins, James T. Linnemann, Jordan Tucker,
Evaluation of three methods for calculating statistical significance
when incorporating a systematic uncertainty into a test of the background-only
hypothesis for a Poisson process, Nuclear Instruments and Methods
in Physics Research A 595 (2008) 480-{}-501.

\bibitem{key-19}G. Cowan, ``Two developments in tests for discovery:
use of weighted Monte Carlo events and an improved measure\textquotedblright ,
Progress on Statistical Issues in Searches\textquotedblright , SLAC'',
June 4 - 6, 2012.

\bibitem{key-20}P. N. Bhattiprolu, S.P. Martin, James D. Wells, Criteria
for projected discovery and exclusion sensitivities of counting experiments,
arXiv:2009.07249 {[}physics.data-an{]}.

\bibitem{key-21}D. Chang, W. Chang, and E. Ma, Alternative interpretation of
the Fermilab Tevatron top events, Phys. Rev. D 59, 091503 (1999. 

\bibitem{key-22}D. Chang, W. Chang, and E. Ma,
Fitting precision electroweak data with exotic heavy quarks,
Phys. Rev. D 61, 037301 (2000).
\end{thebibliography}
\end{document}